\pdfoutput=1
	\newcommand{\apj}{ApJ}
	\newcommand{\apjl}{ApJL}
	\newcommand{\apjs}{ApJS}
	\newcommand{\aj}{AJ}
	\newcommand{\mnras}{MNRAS}
	
	\newcommand{\aap}{A\&A}
\documentclass[fleqn,usenatbib,usegraphicx]{mn2e}
\usepackage{mathptmx}

\usepackage[T1]{fontenc}
\usepackage{ae,aecompl}

\usepackage{graphicx}	
\usepackage{amsmath}	
\usepackage{amssymb}	
\usepackage{subfigure}
\usepackage{url}
\usepackage{deluxetable}
\usepackage{caption}

\title[Stellar Populations in the Tidal Tails of NGC 3256]{A Tale of Two Tails: Exploring Stellar Populations in the Tidal Tails of NGC 3256}

\author[M. A. Rodruck et al.]{
Michael Rodruck,$^{1}$\thanks{E-mail: mrodruck@psu.edu}
Iraklis Konstantopoulos,$^{2}$
Karen Knierman,$^{3}$$^,$$^{4}$
Konstantin Fedotov,$^{5}$$^,$$^{6}$ \newauthor
Brendan Mullan,$^{7}$$^,$$^{8}$
Sarah Gallagher,$^{5}$
Patrick Durrell,$^{9}$
Robin Ciardullo,$^{1}$$^,$$^{10}$ \newauthor
Caryl Gronwall,$^{1}$$^,$$^{10}$ 
and Jane Charlton$^{1}$
\\
$^{1}$Department of Astronomy and Astrophysics, The Pennsylvania State University, University Park, PA 16802, USA\\
$^{2}$Australian Astronomical Observatory, North Ryde, NSW 2133, Australia\\
$^{3}$School of Earth \& Space Exploration, Arizona State University, 550 E. Tyler Mall, Room PSF-686 (P.O. Box 871404), Tempe, AZ 85287-1404, USA\\
$^{4}$NSF Astronomy and Astrophysics Postdoctoral Fellow\\
$^{5}$University of Western Ontario, London, ON N6A 3K7, Canada\\
$^{6}$Herzberg Institute of Astrophysics, National Research Council of Canada, Victoria, BC V9E 2E7, Canada\\
$^{7}$Point Park University Natural Sciences and Engineering Department 201 Wood St Pittsburgh PA 15222, USA\\
$^{8}$Blue Marble Space Institute of Science 1001 4th Ave, Suite 3201 Seattle, WA 98154, USA\\
$^{9}$Department of Physics and Astronomy, Youngstown State University, Youngstown, OH 44555, USA\\
$^{10}$Institute for Gravitation and the Cosmos, The Pennsylvania State University, University Park, PA 16802, USA
}

\date{Accepted XXX. Received YYY; in original form ZZZ}

\pubyear{2016}

\begin{document}
\label{firstpage}
\pagerange{\pageref{firstpage}--\pageref{lastpage}}
\maketitle

\begin{abstract}

We have developed an observing program using deep, multiband imaging to probe the chaotic regions of tidal tails in search of an underlying stellar population, using NGC 3256's 400 Myr twin tidal tails as a case study. These tails have different colours of $u - g = 1.05 \pm 0.07$ and $r - i = 0.13 \pm 0.07$ for NGC 3256W, and $u - g = 1.26 \pm 0.07$ and $r - i = 0.26 \pm 0.07$ for NGC 3256E, indicating different stellar populations. These colours correspond to simple stellar population ages of $288^{+11}_{-54}$ Myr and $841^{+125}_{-157}$ Myr for NGC 3256W and NGC 3256E, respectively, suggesting NGC 3256W's diffuse light is dominated by stars formed after the interaction, while light in NGC 3256E is primarily from stars that originated in the host galaxy. Using a mixed stellar population model, we break our diffuse light into two populations: one at 10 Gyr, representing stars pulled from the host galaxies, and a younger component, whose age is determined by fitting the model to the data. We find similar ages for the young populations of both tails, ($195^{-13}_{+0}$ and $170^{-70}_{+44}$ Myr for NGC 3256W and NGC 3256E, respectively), but a larger percentage of mass in the 10 Gyr population for NGC 3256E ($98^{+1}_{-3}\%$ vs $90^{+5}_{-6}\%$). Additionally, we detect 31 star cluster candidates in NGC 3256W and 19 in NGC 2356E, with median ages of 141 Myr and 91 Myr, respectively. NGC 3256E contains several young (< 10 Myr), low mass objects with strong nebular emission, indicating a small, recent burst of star formation.

\end{abstract}




\section{Introduction}
Galaxy-galaxy interactions lead to the redistribution of stars and gas about each system, and an infusion of material into the local intergalactic
medium, promoting star formation (\citealp{schweizer_87}). Tidal tails are signatures of galactic mergers (\citealp{schweizer_78}), illuminated by the ignition of star formation
(\citealp{Schombert_90}).
Turbulent energy injected into the local H\textsc{i} through these mergers compresses the gas, forming new stars
(\citealp{renaud_14}); This is observationally shown in \cite{mullan_13}, who were able to link
the presence of Star Clusters Candidates (SCCs) to turbulent regions of H\textsc{i}.
While the young, in-situ formed SCCs of tidal tails have been studied in the past through imaging and spectroscopy (e.g. \citealp{knierman_03}; \citealp{trancho_07_1}; \citealp{mello_11}; \citealp{bastian_09}; \citealp{mullan_11}; \citealp{flores_12}), the composition of the underlying stellar material remains a mystery.

We know that gas is easily extracted from the parent galaxies during interactions (e.g. \citealp{hibbard_01}), but whether or not stars follow suit has not yet been established. The extracted gas can collapse to form stars in a clustered manner, yet at the same time, simulations have shown that clusters in tails can be easily disrupted (\citealp{kruij_11}). This sparks an obvious question: what are tidal tails made of? What is the relative fraction of gas and old stars within the interaction ejecta? The answer can inform dynamical simulations of interactions, and therefore help refine our understanding of the enrichment of the intergalactic medium. In fact, current dynamical simulations (e.g. \citealp{renaud_09}) use an older stellar component as a gravitational ‘anchor’ for the gas. Is this requirement justified by the observations? 

Existing studies have focused on the young stellar components of tidal tails, and therefore do not have the capability to probe deep into these regions. We have developed a new observing program designed to do just this, using deep, photometric \textit{ugriz} imaging. To
derive an accurate age estimate for a tidal tail, we plan our exposure times to view across
the stellar sequence in our diffuse tidal tail light. These tails are imaged in each filter to 
an adequate signal to noise ratio, allowing us to derive an average colour and age of their diffuse light. This has the additional benefit of allowing
us to age date star clusters within the tail.

We choose the twin tidal tails of NGC 3256 as a case study for our method. The system is relatively nearby at a distance of 38 Mpc (\citealp{knierman_03}), with an interaction age
of 400 Myr (\citealp{knierman_03}) - not so young that our observations will be drowned out by OB stars, yet young enough so that the tidal structure is still
visible. It has been well studied in the past through
spectroscopic (\citealp{trancho_07_1}; \citealp{trancho_07_2}; \citealp{lipari_00}), photometric (\citealp{mullan_11}; \citealp{knierman_03}; \citealp{zepf_99}) and
H\textsc{i} (\citealp{english_03}) observations, giving us benchmarks to compare to.

We will begin in Section \ref{sec:2} by describing our data and our reduction process. In Section
\ref{sec:3} we describe our analysis methods, and in Section \ref{sec:4} we show our results. In Section
\ref{sec:5} we discuss our findings, and conclude our paper with the main points of our research in
Section \ref{sec:6}.

\section{Data Reduction and Processing}
\label{sec:2}

\subsection{Reduction}

\begin{figure*}
	\centering
	\includegraphics[width=0.5\linewidth]{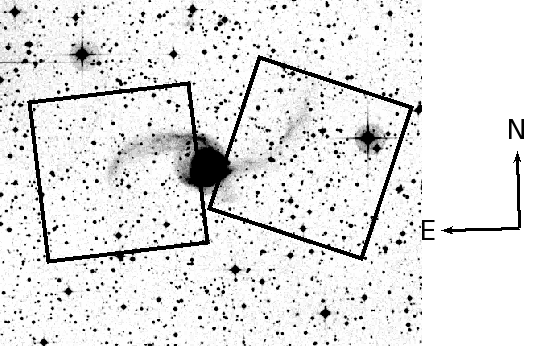}
	\caption{Scanned blue image of NGC 3256 from the Digital Sky Survey. The GMOS field of view is shown in black boxes, and measures $5.6 \times 5.6$ arcmin.}
	\label{fig:GMOS_FOV}
\end{figure*}

Images were obtained from the Gemini-South Observatory from March 2013 to
June 2013, using the GMOS imager. The GMOS field of view is superimposed on an optical image from the Digital Sky Survey in Figure \ref{fig:GMOS_FOV}, measuring
$5.6 \times 5.6$ arcmin. Exposure details are shown in Table \ref{table:exposures}.

\begin{table*}
	\begin{center}
		\begin{tabular}{c c c c c}
			Tail & Filter & Number of exposures & Exposure time per frame (single pointing) & Observation date(s)\\
			& & & (s) & \\ \hline \hline
			West & $u$ & 13 & 600 & March 17, 2013; April 13, 2013 \\  
			West & $g$ & 6 & 150 & May 11, 2013; June 3, 2013\\  
			West & $r$ & 3 & 150 & May 11, 2013 \\  
			West & $i$ & 5 & 200 & May 11, 2013 \\  \hline
			East & $u$ & 13 & 600 & March 12, 2013; March 13, 2013 \\  
			East & $g$ & 3 & 150 & March 17, 2013\\  
			East & $r$ & 3 & 150 & March 17, 2013 \\  
			East & $i$ & 5 & 200 & March 17, 2013 \\  
			\hline
		\end{tabular}
	\end{center}
	\caption{Exposure numbers and times for the Eastern and Western tail. Also listed are dates covering our observations. Data for the $z$-band is excluded (see Section \ref{sec:fringing}).}
	\label{table:exposures}
\end{table*}

\begin{figure*}
	\centering
	\includegraphics[width=1\linewidth]{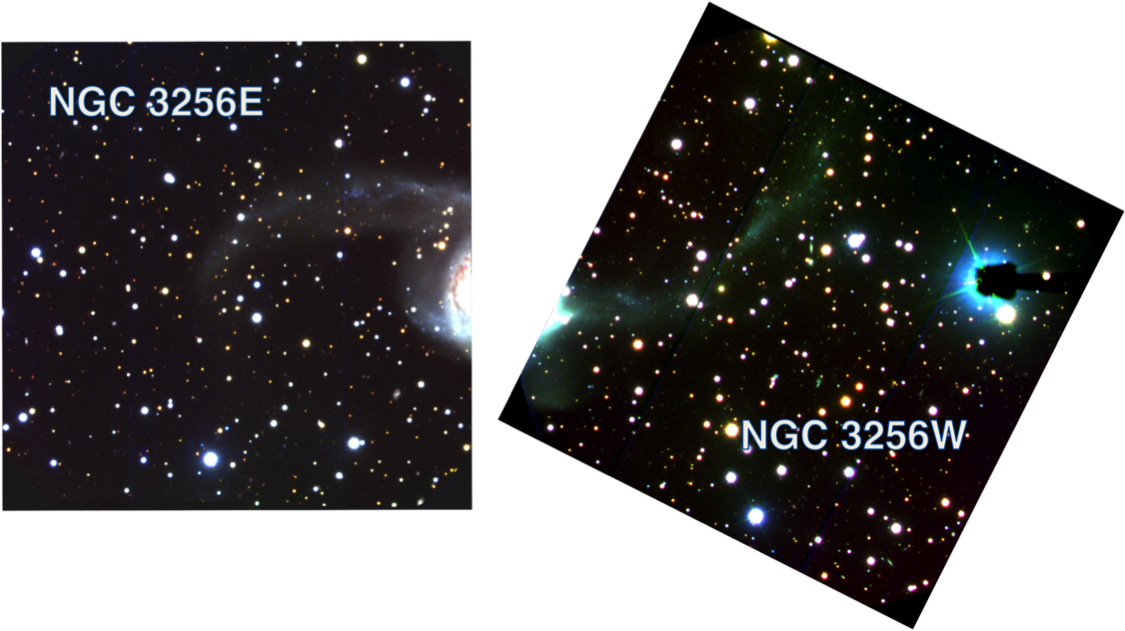}
	\caption{Colour images of NGC 3256E (\textit{left}) and NGC 3256W (\textit{right}) taken with GMOS-S.
		Combined images in the $u$-band represent blue, $g$-band images represent green, and $r$-band
		images represent red. In this image, North is up, and East is to the left.}
	\label{fig:3256_whole}
	
\end{figure*}

Bias and flat field frames were provided by the observatory. Bias
frames were taken every day, but flat fields were taken approximately once
every 2 weeks. The flats closest to the observation date were
used for flat fielding. We used IRAF for data reduction.
Reduced images were mosaiced to obtain a single FITS file from the three CCDs. Combined frames from the \textit{u, g,} and \textit{r}
filters were compiled together in DS9 to create a composite colour image of each tail. Figure \ref{fig:3256_whole}
shows colour frames of NGC 3256E and W. Science frames were calibrated using images of standard star fields (\citealp{smith_02}) taken
each night.

\subsection{Fringing}
\label{sec:fringing}
Fringing is caused by thin-film interference effects in the CCD array
as photons are reflected within the array. This effect is largely
seen in the longer wavelengths, where the optical depth of photons
through the silicon detector is large enough to allow photons to
reflect off the back of the structure and interfere with incoming photons. 
Since the interference is dependent on the thickness of the detector,
the fringe pattern will remain nearly constant, although it can vary
over time in intensity.

\begin{figure*}
	\centering
	\subfigure{
		\scalebox{1}[-1]{\includegraphics[width=0.45\linewidth]{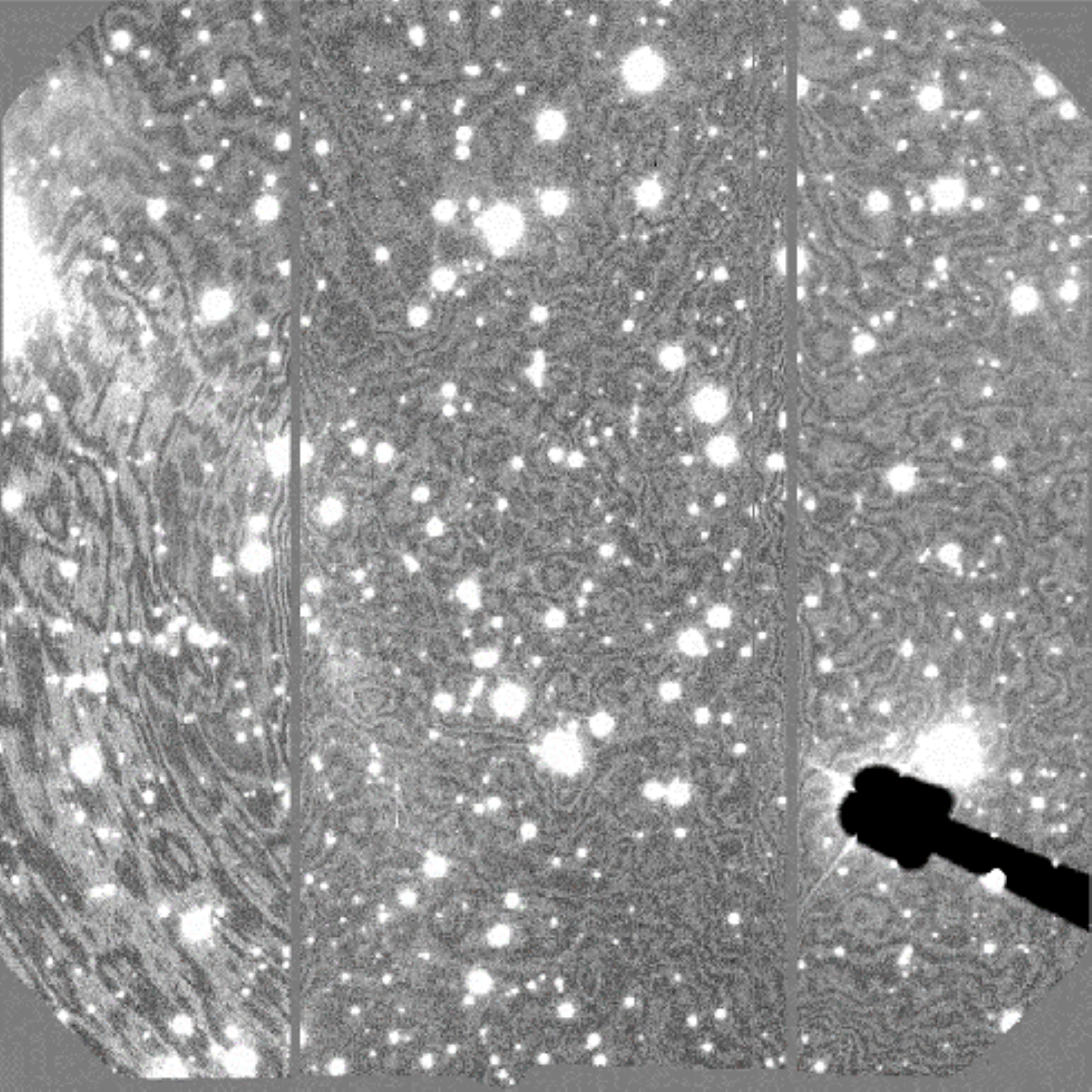}}
	}
	\hspace{1 cm}
	\subfigure{
		\scalebox{1}[-1]{\includegraphics[width=0.45\linewidth]{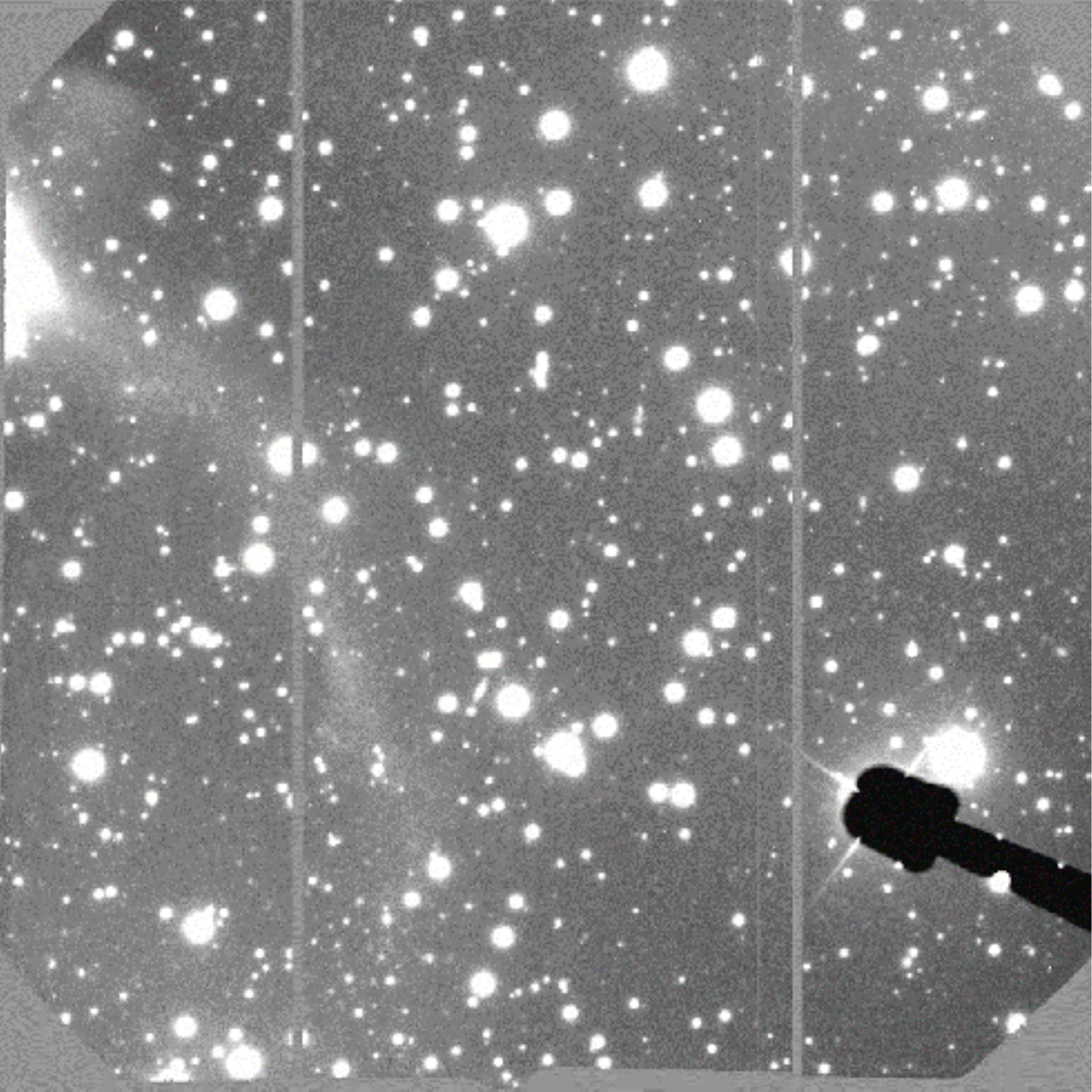}}
	}
	\caption{\textit{Left}: \textit{i}-band image of the Western tail prior to fringe removal. The shadow of the
		telescope's wavefront sensor can be seen in the upper right. \textit{Right}: Same image
		after the fringe pattern has been removed. Note the residual gradient across the image, resulting in bright
		patches at the upper left and lower right corners.}
	\label{fig:Fringe}
\end{figure*}

Both the \textit{i}-band and the \textit{z}-band were severely affected by fringing.
Fringe frames provided by the 
observatory, containing the fringing pattern, allowed for its removal. Unfortunately, the 
fringing in the \textit{z}-band was too severe and could not be corrected. Data in the $z$-band
are therefore not used in our analysis. For the
\textit{i}-band, a median combined frame, compiled from several fringe frames, was needed
to adequately subtract the pattern. Results of the fringe removal for the Western tail
are shown in Figure \ref{fig:Fringe}.

Although the fringing pattern was removed, $i$-band data in the Western tail was left with a slight gradient across the image. This 
is an effect of the subtraction process and imperfect fringe frames, and can affect our measurements of
diffuse light in the \textit{i}-band. Data for the Eastern tail does not show this gradient.

\subsection{Masking Images}
\label{sec:masking}

Effective Low Surface Brightness (LSB) photometry requires the elimination of contaminant
sources. We are attempting to observe the dim, diffuse light of the tidal tail, 
but light from foreground stars and background galaxies can easily dominate. It
is necessary to mask these sources from our images before performing analysis.
Masking the images involves three steps: masking point sources, masking diffuse
sources, and masking bad pixels.

\subsubsection{Magnitude Masking}
To determine how far the influence of a point source object will extend, we create fake stars using \textit{addstar} given a psf generated with the task \textit{psf}. Bright, isolated stars are used to create the template psf. Stars are added to the tail and photometry is performed using the sum of all points inside a list of apertures. The same process is performed on the original image of the tail with no additional stars, and the two sum values are compared to each other for each aperture radius. The radius at which the precision $P$ is less than 0.001 is given as the masking radius, where the precision is defined as:
\begin{equation}
	P=\frac{|\Sigma_{\mathrm{stars}} - \Sigma_{\mathrm{tail}}|}{\Sigma_{\mathrm{tail}}}.
\end{equation}

This radius is plotted against the magnitude of the added stars and a
linear magnitude-radius relationship is found using the IDL procedure
\texttt{linfit.pro}. This process has the weakness that light from
the added
stars cuts off at a radius defined by the parameter \textit{psfrad} in
\textit{daopars}. This makes it impossible to add bright stars ($m \approx 20$) for our analysis, as
the light does not drop below a precision of 0.001
earlier than \textit{psfrad}. Faint stars must be used instead, and we extrapolate
the masking radius to brighter magnitudes, using the linear magnitude-radius relationship.

Objects are found in the image frames using the IRAF task
\textit{daofind}. One frame in each filter is used to find the object
coordinates, which are then used for every other image in the
filter. Photometry is performed using \textit{phot} in IRAF with an
aperture radius of 6 pixels. We apply a mask to objects with magnitudes
less than 25, to protect against spurious detections. This will include objects
identified as SCCs.

Note that this procedure will not work for saturated objects,
as their magnitudes cannot be accurately calculated. The masking will
be underestimated, and a manual radius will need to be set. Fortunately,
this was only needed for 7 objects in the Western tail, and none in the Eastern.

\subsubsection{Object Masking}
The IRAF task \textit{objmask} will find objects in an image which are
$n \cdot \sigma$ above background. This is useful for detecting
cosmic rays and non point source objects (e.g., resolved galaxies). It is set to detect objects which are a minimum of
2 pixels in size to account for cosmic rays.

\subsubsection{Bad Pixel Masks}
The Gemini IRAF package includes a set of pixel masks for detectors at
both the North and South sites. However, streaks in the images appear 
in several consistent locations which are not covered by the pixel masks. 
A new mask is created with manual corrections for these streaks. This mask
has the same shifts applied to it as the aligned images, resulting in a 
unique bad pixel mask for every image.

\begin{figure*}
	\centering
	\subfigure{
		\scalebox{1}[-1]{\includegraphics[width=0.45\linewidth,height=0.45\linewidth]{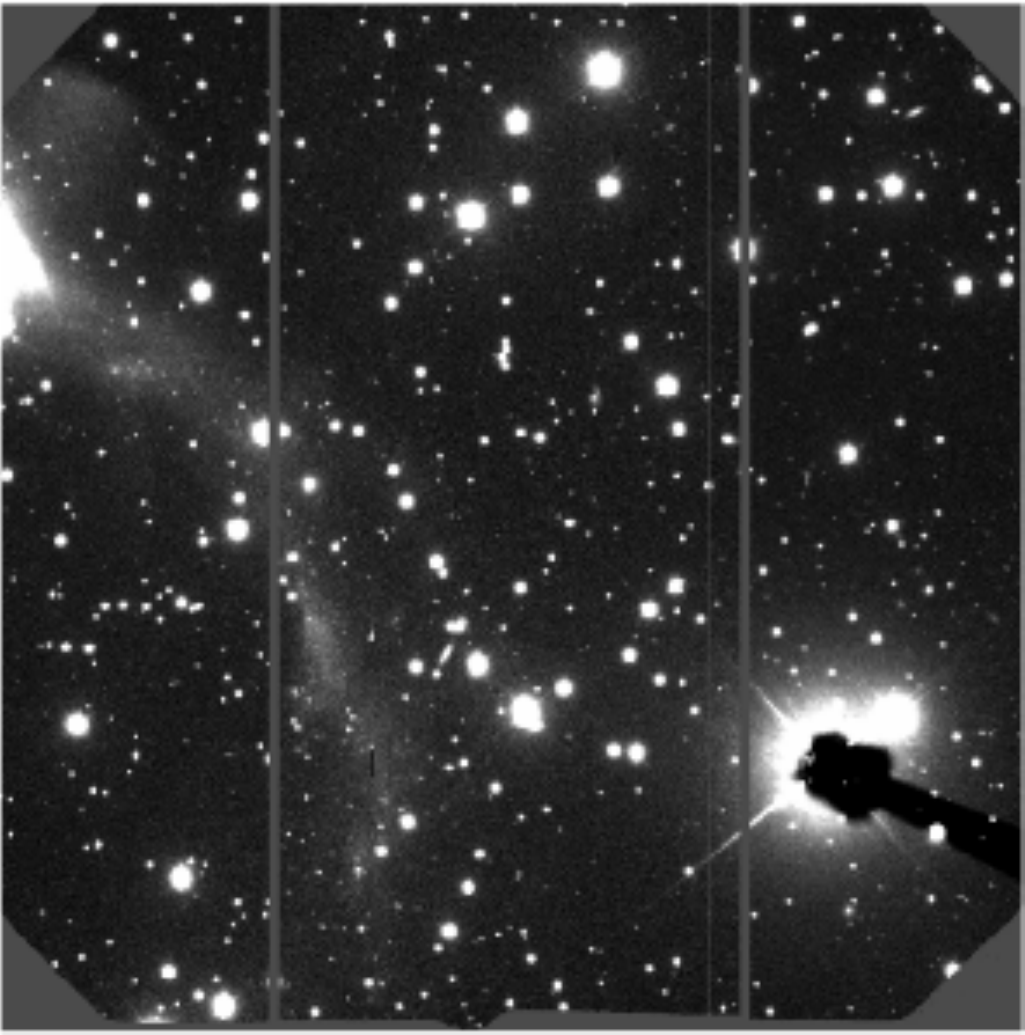}}
	}
	\hspace{1 cm}
	\subfigure{
		\scalebox{1}[-1]{\includegraphics[width=0.45\linewidth,height=0.45\linewidth]{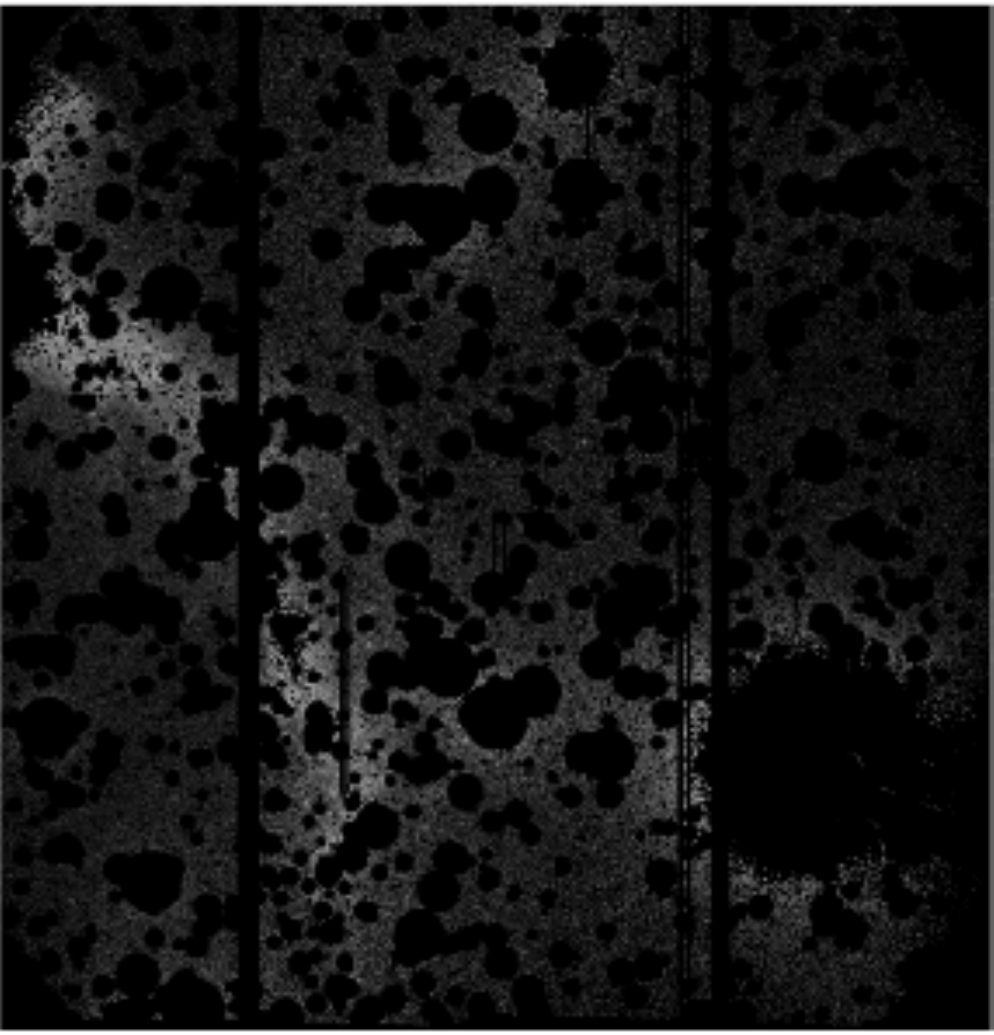}}
	}
	\caption{\textit{Left}: \textit{g}-band image of NGC 3256W. \textit{Right}: Same image, but with the
		mask applied. The tail light stands out free of contaminant sources. Note the diffuse light from
		several brighter stars appears in the top and bottom of the image. These objects were saturated and an 
		accurate mask could not be found. However, they are far enough from the tail to not have an effect.}
	\label{fig:West} 
	\vspace{0.5 cm}
\end{figure*}

The three masks are compiled into one composite image which is multiplied 
by every science frame. An example of this process is shown in Figure \ref{fig:West}.

\section{Data Analysis}
\label{sec:3}

\subsection{Low Surface Brightness Photometry}
Use of the image masks will eliminate contaminant sources, leaving
the diffuse light of the tail untouched. The background level of the 
sky must be accurately measured and subtracted from the tail to find its 
surface brightness. 

\subsubsection{Background determination}
The GMOS detector consists of three E2V CCDs. Each of these chips has a 
background level differing from the others. We sample 25 unique locations
on each chip with square apertures of dimensions 41 $\times$ 41 pixels, or 6 $\times$ 6 arcseconds. Locations
are selected to be near the tail to achieve representative sampling.

We fit the data to a Gaussian distribution using a maximum loglikelihood estimator. 
Initial guesses at the mean and standard deviation values are found by
plotting the data in a histogram and fitting the binned points to a 
Gaussian curve using the IDL program \texttt{gaussfit}. As this method
is sensitive to the size of the histogram bins, we use this as first approximation
and perform our final analysis with the more robust maximum loglikelihood estimator.

\subsubsection{Tail Definition} 
\label{sec:tail_def}

Images in the \textit{g}-band were combined for defining tail regions using \textit{imcombine}, prior to masking. We chose \textit{g}-band data as
it offered the best seeing and a good signal to noise. The combined frames were then smoothed with a Gaussian
kernal with a FWHM of 5-6 pixels, in order to boost the S/N in each pixel. Regions defined as ``in-tail'' were determined as contiguous regions 3$\sigma$
above the background, using the IDL program \texttt{find\_boundary.pro}. Average background and standard
deviation values were determined in each chip from the co-added image. Regions between adjacent chips were manually added
to extend the tail across the CCDs. 

For the Eastern tail, we manually cut the region to exclude the bulge and southern tidal debris, so as to focus
on the tail itself. Another cut was made at the tip of the tail to exclude regions marked by diffuse light from
a nearby bright star. The Western tail also includes a manual cut on the south-eastern edge to exclude an extended region due to diffuse light
from bright stars. Figure \ref{fig:Boxes} contains our boundaries considered for photometry. Observations for the Western tail include a diffuse feature seen at the bottom of the image, separate from
the tail (see Figure \ref{fig:Boxes}). We include measurements for this region, but consider it separately from the tail itself.

\subsubsection{Photometry}
\begin{figure*}
	\centering
	\subfigure{
		\centering
		\includegraphics[width=0.45\linewidth]{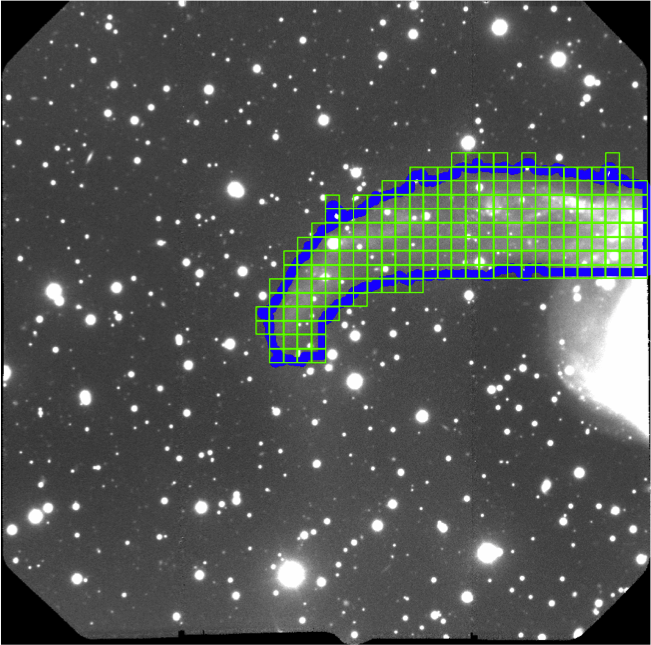}
	}
	\hspace{1 cm}
	\subfigure{
		\includegraphics[width=0.45\linewidth]{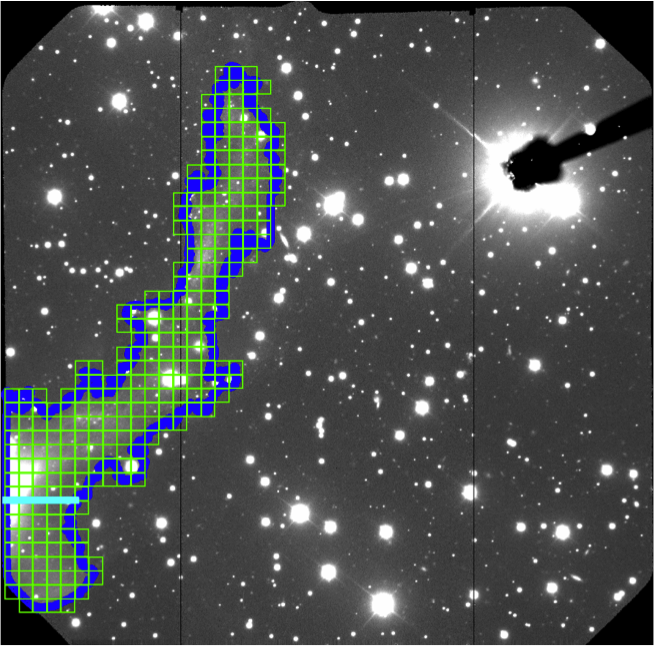}
	}
	\caption{Selected regions for photometry for NGC 3256E (\textit{left}) and NGC 3256W (\textit{right}). Blue
		lines indicate the boundary of the tail region. In the Western tail, a cyan line indicates the boundary between the tail and diffuse structure. Several boxes are not fully enclosed with the tail, however
		only pixels within the blue boundary are counted for photometry. We additionally only count boxes with $\geq$ 50
		good pixels.}
	\label{fig:Boxes}
	
\end{figure*}

We apply a grid of boxes 50 $\times$ 50 pixels in size to both tails for our measurement apertures of diffuse
light, as show in Figure \ref{fig:Boxes}, allowing us to determine the spatial distribution of colours in the tail. Box sizes correspond to 7.3 $\times$ 7.3 arcseconds on the sky, or 365 $\times$ 365 pc on the tail. The tail boundaries are shown in blue. Several boxes are not fully bounded inside the blue tail boundary, however, only the pixels within the tail in these boxes are counted for photometry. We limit analysis to boxes with $\geq 50$ pixels within the tail. For NGC 3256W we mark
the boundary between the diffuse structure and tail itself in cyan.

We apply a 3$\sigma$ limit to our detections. The flux in each measurement box was corrected for airmass and Galactic dust extinction.
Foreground extinction corrections were obtained from \cite{schlafly_11}, using an $R_\lambda = A_\lambda/E(B-V)$ reddening law, with $R_V = 3.1$.
The corrected flux was averaged across images in the filter and converted to a surface brightness
in units of mag $\cdot$ arcsec$^{-2}$. Magnitude zero points and airmass correction
coefficients were taken from the Gemini Observatory webpage\footnote{\url{https://www.gemini.edu/?q=node/10445}}, and are listed in Table \ref{table:zpoints}.

\begin{table}
	\begin{center}
		\begin{tabular}{c c c}
			Filter & Magnitude zero point & Airmass correction coefficient\\ \hline \hline
			\textit{u} & 24.91 & 0.38 \\  
			\textit{g} & 28.33 & 0.18 \\  
			\textit{r} & 28.33 & 0.10 \\  
			\textit{i} & 27.93 & 0.08 \\
			\hline
		\end{tabular}
	\end{center}
	\caption{Magnitude zero points and airmass correction coefficients for the E2V CCDs}
	\label{table:zpoints}
\end{table}

\subsection{Cluster Photometry}
NGC 3256 is far enough away, at 38 Mpc, that its clusters will not be resolved from the ground, and will appear
as point sources. For a typical seeing of 0.9", objects less than 74 pc will appear unresolved. Thus, our photometry will
be performed on these unresolved, point sources.

\subsubsection{Aperture Corrections}

The imaging field is crowded enough to require aperture corrections to prevent
contamination from outside objects. A curve of growth analysis using 10 bright stars
was performed for each filter. Given the varying PSF of each image, it was necessary to 
extend this analysis to every image, so that each image has its own aperture correction.
We corrected for a measurement radius of 6 pixels, for every image. For reference, the FWHM
of our images ranged from $\sim$4.5 pixels, to $\sim$9.5 pixels, depending on the seeing.
\subsubsection{Photometry}
\begin{figure*}
	\centering
	\subfigure{
		\centering
		\includegraphics[width=0.45\linewidth,height=0.45\linewidth]{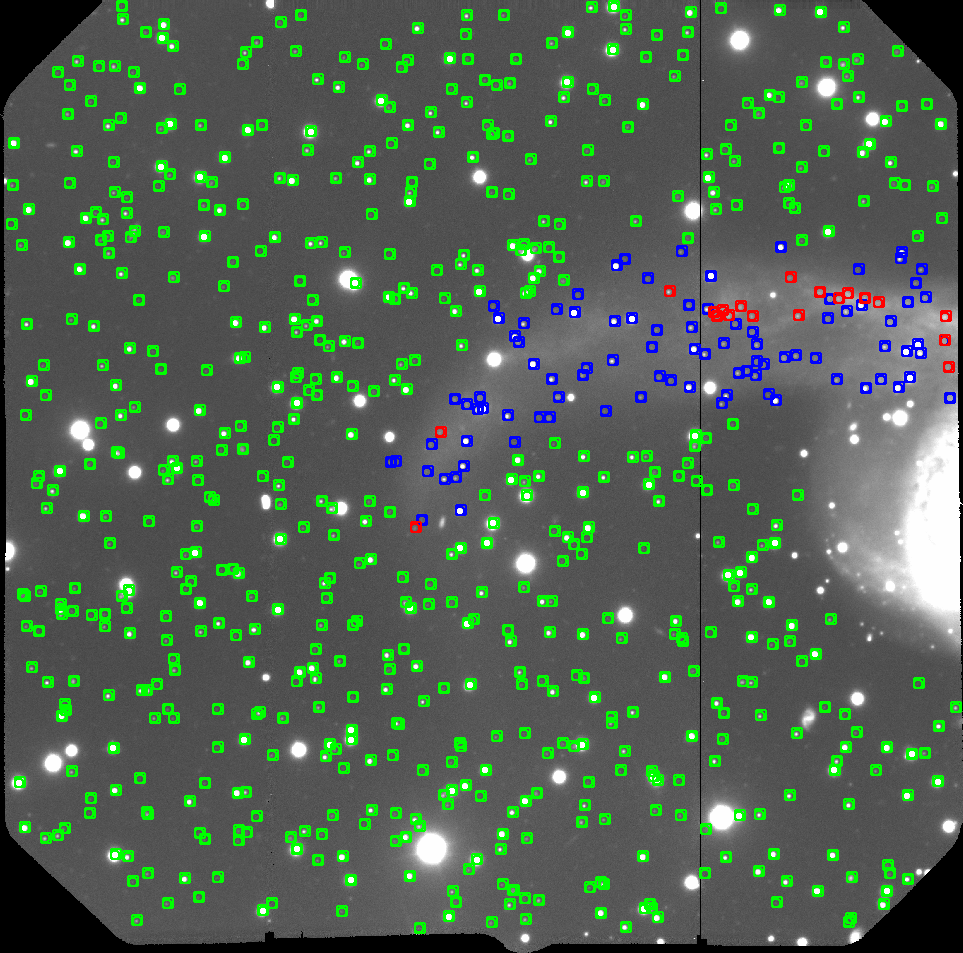}
	}
	\hspace{1 cm}
	\subfigure{
		\includegraphics[width=0.45\linewidth,height=0.45\linewidth]{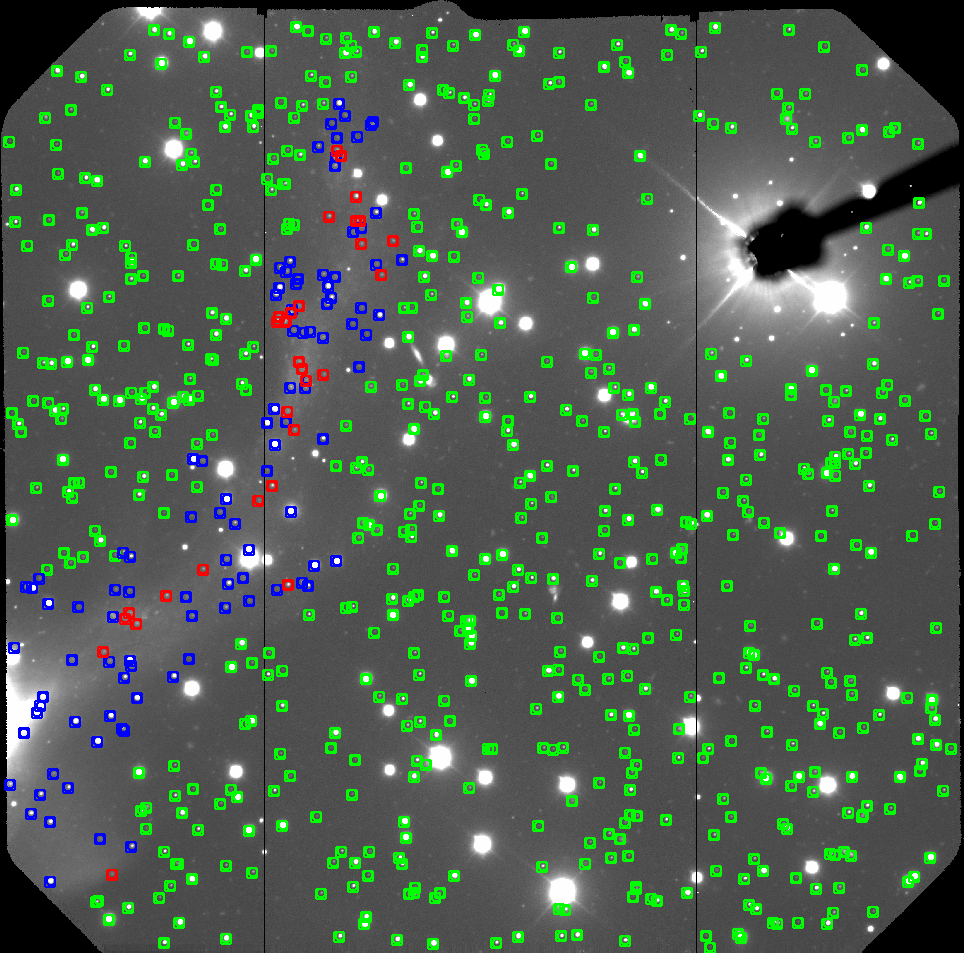}
	}
	\caption{\textit{Left}: Locations for in-tail objects (blue) and out-of-tail objects (green)
		shown for NGC 3256 E. Red sources are Star Cluster Candidates (SCCs), as determined in Section \ref{sec:cluster_colours}. \textit{Right}: Same, but for the Western tail.}
	\label{fig:Objs}
	
\end{figure*}

Objects were detected in \textit{g}-band combined images using the IRAF task \textit{daofind}. Sources were labeled 
as ``in-tail'' or ``out-of-tail'' using boundaries as defined in section \ref{sec:tail_def}. Detections that were
cosmic rays, saturated stars, resolved galaxies, or duplicate detections were manually removed. Figure \ref{fig:Objs} shows
locations of detected objects.

We used the IRAF task \textit{phot} to perform aperture photometry on our detected objects, with an aperture size of 6 pixels. We considered sources > 3$\sigma$ above the background. To remove cosmic rays
in our frames, we used the IDL routine \texttt{la\_cosmic} on our dataset. Objects containing cosmic rays in their
measurement aperture were discarded for that particular frame. 


\section{Results}
\label{sec:4}
\subsection{Tidal Tail Colours}
\label{section:LSB_colours}
\begin{figure}
	\centering
	\includegraphics[width=1\linewidth]{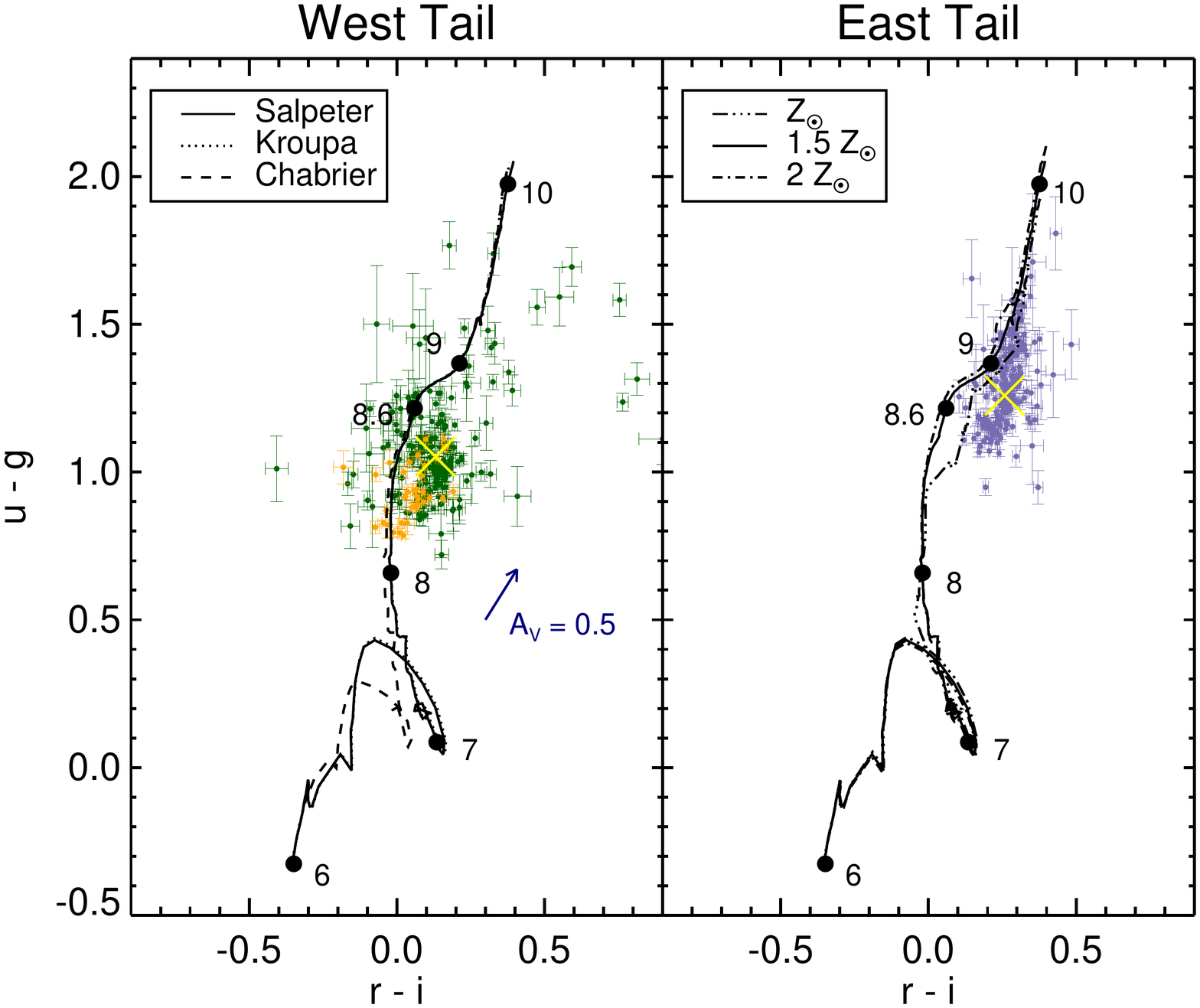}
	\vspace{-0.5cm}
	\caption{Colour-colour plot of NGC 3256W (\textit{left}) and E (\textit{right}). Points correspond to integrated
		light from green boxes in Figure \ref{fig:Boxes}. Green and purple points mark data from the Western and
                Eastern tails respectively, while orange points for NGC 3256 indicate data from the separate diffuse
                structure. Numbers mark the logarithmic age of the model from 1 Myr to 10 Gyr. A yellow ``X'' marks
		each tail's overall colour. In each
		plot we include a 1.5 $Z_{\odot}$ SSP evolutionary track with a Salpeter IMF as a solid black line. On the left, we
		include tracks using Kroupa and Chabrier IMFs, at 1.5 $Z_{\odot}$, to highlight the insensitivity of the modeled colours to the
		choice of IMF. On the right, we include tracks with a range of metallicities to again highlight the minimal effect of metallicity on colour.}
	\label{fig:LSBphot}
	
\end{figure}

Integrated surface brightness, colour, and age measurements are included in Table \ref{table:LSBmeas}. We plot
colours for the West and East tails in Figure \ref{fig:LSBphot}. An extinction vector of A$_V$ = 0.5 is shown. Data for regions within the diffuse structure
in the Western tail are shown in orange, while measurements from regions in the tail itself are shown in green. In the right panel, the colours of the different regions of the Eastern tail are shown in purple. The
overall tail colours from Table \ref{table:LSBmeas} are marked by yellow ``X's''. We plot these
colours against Simple Stellar Population (SSP) evolutionary models from \cite{marigo_08} for a metallicity of 1.5 $Z_{\odot}$ and a
Salpeter (\citealp{salpeter_55}) IMF, in solid black. Our choice in evolutionary models follows \cite{iraklis_12}, which found good
fits between \cite{marigo_08} models and their data, for young and old clusters. Our motivation for this metallicity comes from \cite{trancho_07_1}, who 
spectroscopically studied a number of in situ formed star clusters in NGC 3256 and found a mean metallicity of 1.5 $Z_{\odot}$.
Overlaid on data for the Western tail are tracks for a Kroupa and Chabrier IMF at 1.5 $Z_{\odot}$. For
the Eastern tail, we show evolutionary tracks of 1, 1.5, and 2 $Z_{\odot}$ using a Salpeter IMF. The similarity of 
evolutionary tracks using different IMFs and metallicities suggest these will have little effect on our analysis.

\begin{table*}
	\begin{center}
		\resizebox{\textwidth}{!}{
		\begin{tabular}{c c c c c c c c c}
			Tail & \textit{u} & \textit{g} & \textit{r} & \textit{i} &  $u - g$ & $r - i$ & Diffuse Light Age & Interaction Age$^1$   \\
			& (mag arcsec$^{-2}$) & (mag arcsec$^{-2}$) & (mag arcsec$^{-2}$) & (mag arcsec$^{-2}$) &  & & (Myr) & (Myr) \\  \hline \hline
			West & $25.59 \pm 0.05$ & $24.54 \pm 0.05$ & $24.22 \pm 0.05$ & $24.09 \pm 0.05$ & $1.05 \pm 0.07$ & $0.13 \pm 0.07$ & $288^{+11}_{-54}$ & 400 \\[0.15cm]  
			East & $25.49 \pm 0.05$ & $24.23 \pm 0.05$ & $23.77 \pm 0.05$ & $23.51 \pm 0.05$ & $1.26 \pm 0.07$ & $0.26 \pm 0.07$ & $841^{+125}_{-157}$ & 400 \\
			\hline
		\end{tabular}}
	\end{center}
	\caption{Measured properties of NGC 3256. Note that these values are \textit{not} averages, but are instead found by integrating the tail light with the boundaries shown in Figure \ref{fig:Boxes} as a whole. Photometry errors are negligible, due to sampling the large area of the tidal tail.
		We estimate systematic errors at a minimum of $\pm 0.05$ mag.\newline $^1$From \protect \cite{knierman_03}}
	\label{table:LSBmeas}
\end{table*}

\begin{figure}
	\centering
	\includegraphics[width=1\linewidth]{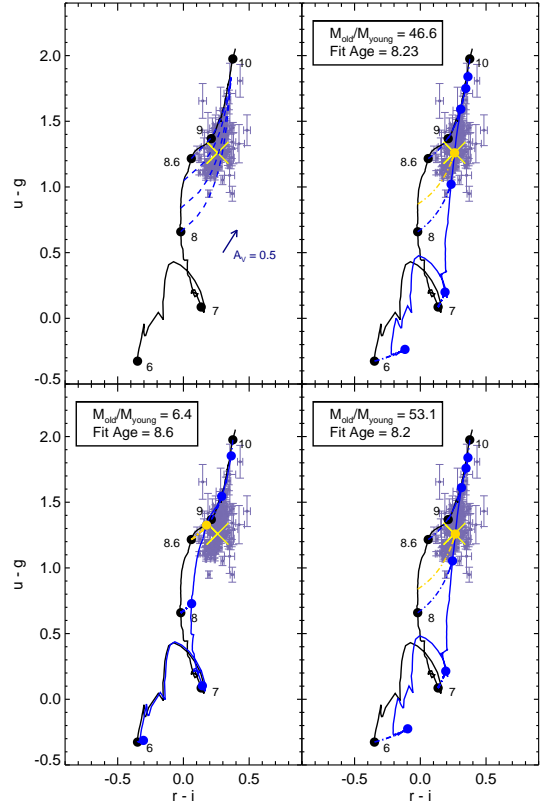}
	\vspace{-0.5cm}
	\caption{\textit{Upper Left}: Dashed blue lines
		indicate the shift in evolutionary track colours for specific young population ages (8.0, 8.2, 8.4, 8.6 log yrs), as the ratio of $M_{\mathrm{old}}$/$M_{\mathrm{young}}$ increases, where the old population is a 10 Gyr stellar population. \textit{Upper Right}: Best fit MSP evolutionary track
		(blue)
		to overall tail	colours, at a mass ratio of $M_{\mathrm{old}}$/$M_{\mathrm{young}} = 46.6$. Gold circle marks the closest point between the track and the tail colour, corresponding to an age of 8.23 log yrs. \textit{Bottom}: MSP fits to tail colour at
		specified logarithmic ages. The mass ratios are adjusted to find the minimum distance between
		the tail colour and the MSP track colour at the indicated young population age (8.6 and 8.2 log yrs). Gold circle indicates the point on the MSP track corresponding to the age which is being fit.}
	\label{fig:E_age_tracks}
\end{figure}

Results for the Eastern tail are in good agreement with measurements from \cite{mulia_15}, which studied the Eastern tail of NGC 3256, among other tails. A comparison of ages in Table \ref{table:LSBmeas} shows the integrated diffuse light in the Eastern tail is more than twice as old as the 
interaction age of the galaxies, suggesting the stellar light in the Eastern tail is dominated by an older stellar
population, likely originating from the host galaxies. To investigate this, we add light to our SSP model from a stellar population at 10 Gyr
and solar metallicity, creating a Mixed Stellar Population (MSP) model. The influence of the old population reddens both the $u - g$ and $r - i$ colours. We choose SSP models with 4 logarithmic ages (8.0, 8.2, 8.4, and 8.6 log yrs) to demonstrate
this effect, by plotting their progression towards redder colours as we increase the mass ratio between the 10 Gyr
population and this original SSP (see top left of Figure \ref{fig:E_age_tracks}).


In the top right of Figure \ref{fig:E_age_tracks} we fit an MSP model to the tail's integrated colour. The ratio of masses between
the populations is adjusted to minimize the distance between the MSP track and the integrated tail colour.
We can then find the age of the
original SSP model corresponding to these colours to obtain a representative age of the young population. We find the best fit parameters
give a young population at an age of $8.23^{-0.23}_{+0.11}$ log yrs and a mass ratio $M_\mathrm{old}$/$M_\mathrm{young}$ of $46.6^{+87.1}_{-25.8}$, indicating that $98^{+1}_{-3}$\% of the stellar mass and
$68^{+11}_{-13}$\% of the bolometric flux is from a 10 Gyr population. Upper and lower limits are found by fitting our MSP track to the upper and lower
limits of the integrated $u - g$ and $r - i$ colours. This can result in smaller upper limits and larger lower limits for our mass and flux ratios, as the older
ages require less of a 10 Gyr population to shift the track in the appropriate manner. If instead of a 10 Gyr population we included an 8 Gyr population, the mass ratio $M_\mathrm{old}$/$M_\mathrm{young}$ would increase to 
48.7, while a 12 Gyr population would increase $M_\mathrm{old}$/$M_\mathrm{young}$ to 47.5. Note that while the mean colours of the tail are fit to a specific 
population, several regions in the tail can be fit to younger ages at varying mass ratios. In the bottom panels of Figure \ref{fig:E_age_tracks} we include MSP models with young population ages of 8.2 and 8.6 log yrs. Our fit to 8.6 log yrs represents the oldest possible age for new star formation, as stars at this age would have formed immediately after the interaction. This also gave us the minimum mass ratio between the populations as $M_\mathrm{old}$/$M_\mathrm{young}$ = $6.4^{+3.8}_{-2.9}$. Although
the exact numbers vary with location in the tail, and depend on assumptions about the ages of the young
and old populations, we have demonstrated that there is a significant contribution from an old stellar
population to the diffuse light in the Eastern tail.

Our fits to the MSP models assume A$_V$ = 0. We do not expect the role of extinction to be large, as discussed
at the end of Section \ref{sec:5}. We note that the effect of dust would be to decrease the age and mass ratio
$M_\mathrm{old}$/$M_\mathrm{young}$ of the tail, as the data would shift closer to the SSP track, toward bluer colours.

\begin{figure}
	\centering
	\includegraphics[width=1\linewidth]{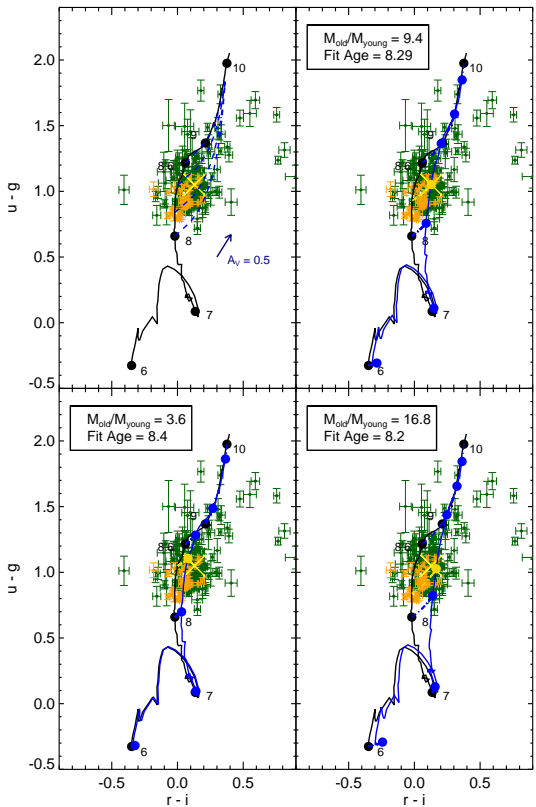}
	\vspace{-0.5cm}
	\caption{Same as Figure \ref{fig:E_age_tracks} but for the Western tail.}
	\label{fig:W_age_tracks}
\end{figure}

We perform the same analysis on the Western tail in Figure \ref{fig:W_age_tracks}. The best fit mass ratio is $M_\mathrm{old}$/$M_\mathrm{young} = 9.4^{+8.1}_{-4.2}$, with an age of $8.29^{-0.03}_{+0.0}$ log yrs. 
However, photometric errors in the \textit{i}-band can inflate the mass ratio between the old
and young populations. The effect of the old population on the $r - i$ colour is to push data points toward 
redder colours. Unfortunately, this same effect could also be caused by photometric scatter from imperfect fringe 
frame removal. As noted in Section \ref{sec:fringing},
fringes in the \textit{i}-band were removed, but a gradient was left across the images. This gradient could have
artificially reddened our $r - i$ colours, which would increase the mass ratio $M_\mathrm{old}$/$M_\mathrm{young}$,
creating a larger old population than truly exists. We note this has a marginal effect on the \textit{age} of the light
(at $8.29^{-0.03}_{+0.0}$ log yrs), as the $u - g$ colour is primarily used to determine the age,
which is clearly dominated by the younger population in this tail.

%
%

\subsection{Tail Stellar Mass}
The stellar mass of each tail region can be calculated using its observed $M_\mathrm{old}$/$M_\mathrm{young}$ ratio and the magnitude in each filter. We are over-constrained, as we will have a unique mass for each filter's magnitude. We take the mean value between filters to use as our mass, and repeat this process for each measurement box shown
in Figure \ref{fig:Boxes}. Each data point is broken into a composite population, one part consisting of a 10 Gyr population, and the second a younger
population, whose age and mass ratio relative to the old population is determined by the MSP fit to the data point. Upper and lower bounds on our masses are found by fitting our
MSP model to the upper and lower error bounds of each data point. This can create smaller upper bounds and larger lower bounds than the nominal values, as the best fit model may require
different mass ratios between the two populations. We compare the masses of each tail to previous measurements of H\textsc{i}, and our observed total cluster
mass as derived in the ensuing section, in Table \ref{table:mass_all}. We use a Salpeter IMF to derive our mass values. While
the choice of IMF does not affect the colours of our objects (and therefore our age derivation), it can have an appreciable
effect on the derived masses. A Chabrier (\citealp{chabrier_01}) or Kroupa (\citealp{kroupa_01}) IMF results in a decrease in mass by a factor of $\sim$2 relative to a Salpeter IMF. 

Our masking of contaminant sources has eliminated a substantial fraction of each tail. In the Western
tail, unmasked pixels represent 41\% of the total tail area; for the Eastern tail, we have 54\% of the
tail left unmasked. We correct our masses of diffuse light in Table \ref{table:mass_all} to account
for the missing tidal tail light.

\begin{table*}
	\begin{center}\resizebox{\textwidth}{!}{
		\begin{tabular}{c c c c c c c c}
			Tail & Diffuse Old  & Diffuse Young  & Median Cluster Age & Diffuse Old Mass & Diffuse Young Mass & Star Cluster Mass & H\textsc{i} Mass$^1$ \\
			&  Age (log yrs) &  Age (log yrs) & (log yrs) & ($10^8 M_{\odot}$) & ($10^8 M_{\odot}$) & ($10^8 M_{\odot}$) & ($10^8 M_{\odot}$) \\ \hline \hline
			West & 10 & $8.29^{-0.03}_{+0.0}$ & 8.15 & $128.3^{+5.6}_{-5.6}$ & $15.1^{+0.5}_{-0.5}$ & $0.31^{+0.05}_{-0.06}$ & 22.0 \\ [0.15cm]
			East & 10 & $8.23^{-0.23}_{+0.11}$ & 7.96 & $392.2^{-2.8}_{+6.1}$ & $10.7^{+1.3}_{-1.3}$ & $0.13^{+0.04}_{-0.02}$ & 14.0 \\
			\hline
		\end{tabular}}
	\end{center}
	\caption{Mass measurements of NGC 3256, using a Salpeter IMF for stellar mass. \hspace{\linewidth} \newline $^1$From \protect \cite{knierman_03}}
	\label{table:mass_all}
\end{table*}


Several data points younger than the interaction age of 8.6 log yrs in the Western tail fall directly on the SSP track. In these cases our colour-colour diagram indicates the mass is completely derived from a young
population, with negligible contribution from an old population. More likely, the presence of the old population is drowned out by the flux
of the young population. To find the maximum old population allowed in such cases, we add the maximum allowed mass such that the
derived magnitude fits within each filters observed magnitude error bars. This marginally increases the overall mass of the tail by an additional 0.2\%. 

\subsection{Star Clusters}

\subsubsection{Cluster Colours}
\label{sec:cluster_colours}

\begin{figure}
	\centering
	\includegraphics[width=1\linewidth]{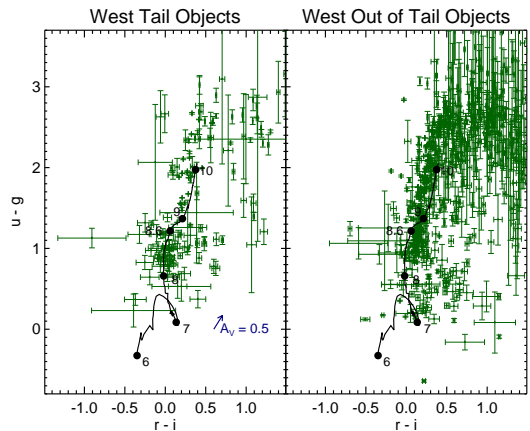}
	\vspace{-0.5cm}
	\caption{Objects detected by \textit{daofind} both in the Western tail (\textit{left}) and outside the tail (\textit{right}).
		We zoom out in this plot and decrease the plot symbol size to highlight the wide colour range of objects detected.}
	\label{fig:Wclusters}
\end{figure}

\begin{figure}
	\centering
	\includegraphics[width=1\linewidth]{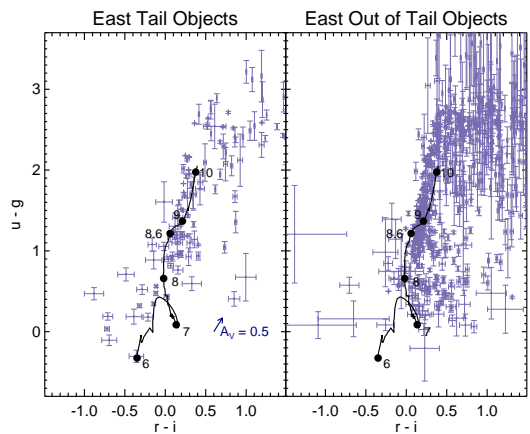}
	\vspace{-0.5cm}
	\caption{Same as Figure \ref{fig:Wclusters} but for the Eastern tail.}
	\label{fig:Eclusters}
\end{figure}

Colour-colour diagrams for point sources detected in the Western and Eastern tails are shown in Figures \ref{fig:Wclusters} and \ref{fig:Eclusters},
respectively. We compare the distribution of sources both in and out of the tail. 

\begin{figure*}
	\centering
	\includegraphics[width=1\linewidth]{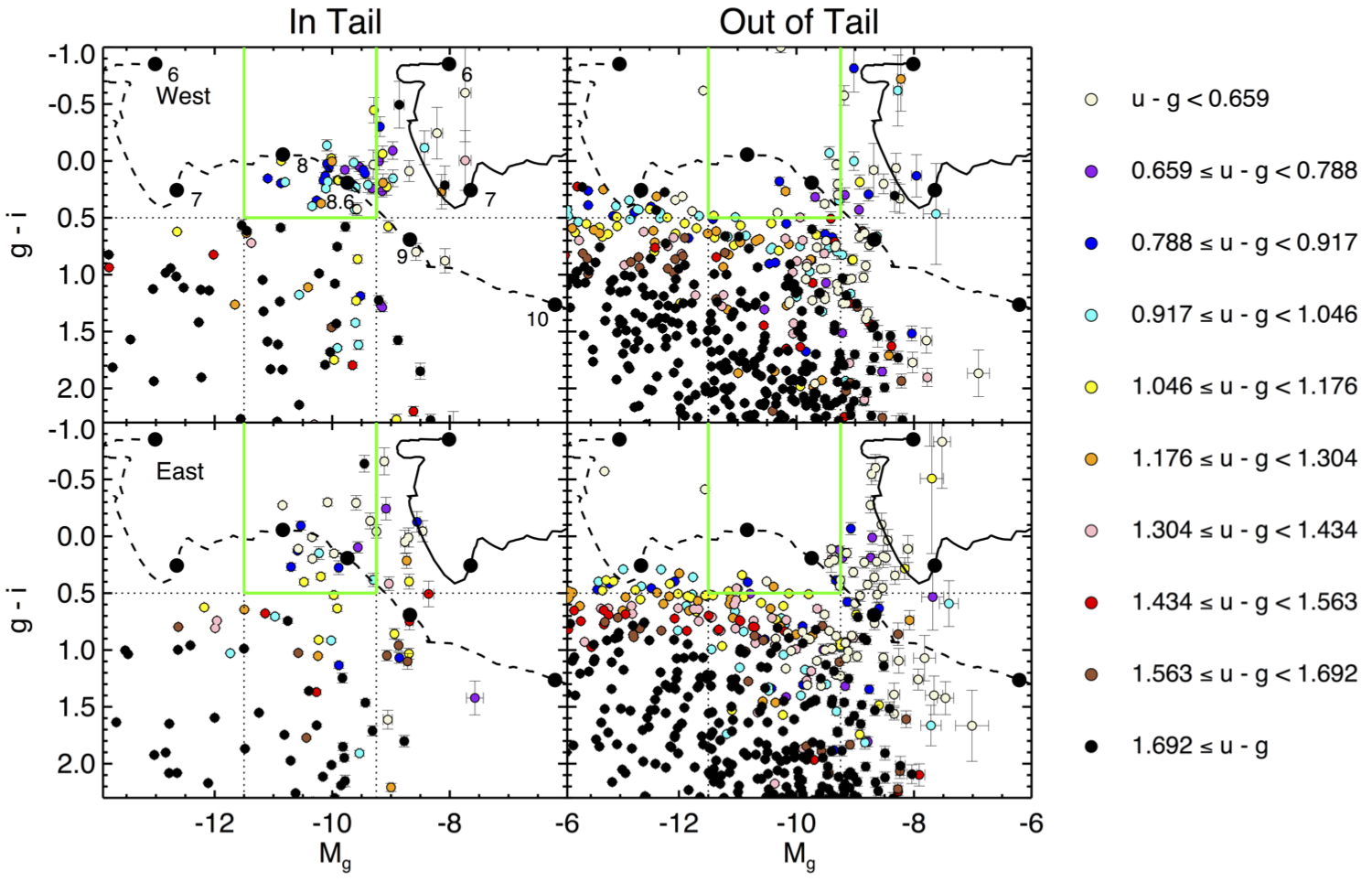}
	\vspace{-0.5cm}
	\caption{Colour-magnitude diagrams for the tidal tails. \textit{Upper}: Data for the Western tail. Left hand panel shows sources
		in the tail, while objects outside of the tail are shown on the right. Colours of sources indicate their $u - g$ values.
		Overlaid are evolutionary tracks for $10^4$ (solid) and $10^6$ (dashed) $M_{\odot}$.
		Dotted lines indicated magnitude and colour cuts. Objects considered to be SCCs lie within the green box. \textit{Lower}: Same, but for the Eastern tail.}
	\label{fig:cluster_masses}
\end{figure*}

Locations for detected sources are shown in Figure \ref{fig:Objs}. Following the above standard, we superimpose a Salpeter
IMF at 1.5 $Z_{\odot}$. An over density of objects can be seen in both tails (see Figures \ref{fig:Wclusters} and \ref{fig:Eclusters}) as compared to their out-of-tail sources, at and
below the 8.6 log yrs marker, suggesting that many of these objects are real clusters. We can see the Western tail has more of these objects, although
the Eastern tail has a grouping of several very young objects. However, in-tail and out-of-tail sources for both tails greatly overlap. To aid in discriminating between contaminating and SCC sources in our tails, we use $g - i$ and $M_g$
cuts as in \cite{knierman_03} and \cite{mullan_11}. We plot colour-magnitude diagrams in Figure \ref{fig:cluster_masses} overlaid with SSP models using a Salpeter IMF for masses of 10$^4$ and 10$^6$ $M_{\odot}$. Colours of data points
indicate their $u - g$ values. 

Both of the out-of-tail plots show an excess of sources greater than $g - i$ = 0.5. A second clustering 
on the faint end is seen at $g - i$ $<$ 0.5 and $M_g < -9.25$. A final cut at $M_g < -11.5$ is made to exclude brighter contaminating sources. 
We thus consider objects with $g - i$ $\leq$ 0.5 and $-11.5 \leq M_g \leq -9.25$ to be 
star cluster candidates (SCCs). 

We present the number of SCCs detected within ($N^\mathrm{SCC}_\mathrm{in}$) and outside ($N^\text{SCC}_\text{out}$) the tail, as well as the areas of each region ($A_\text{in}$ and $A_\text{out}$) in 
Table \ref{table:excess}. In-tail SCC excess $\Sigma_\mathrm{SCC}$ is defined as the difference between 
$N^\text{SCC}_\text{in}/A_\text{in}$ and $N^\text{SCC}_\text{out}/A_\text{out}$. Poisson errors are given
at 1$\sigma$, recorded from $N^\text{SCC}_\text{in}$ and $N^\text{SCC}_\text{out}$. SCC excess in the 
Western tail is significant at $> 5\sigma$, while the Eastern tail excess is significant at $4\sigma$, suggesting most of these objects are real clusters. Our results for the Western tail agree with \cite{mullan_11}, 
which found a statistically significant excess, although at a lower confidence level. However, they found no
significant excess in the Eastern tail. Our values differ mainly for two reasons: our colour and magnitude
cuts are more conservative, and our sample area for the out-of-tail region is larger. \cite{mullan_11} used a colour-magnitude criteria of $V - I < 2.0$ ($g - i \approx 2.2$, determined through fitting \citealp{marigo_08} SSP models) and 
$M_V < -8.5$ ($M_g \approx -8.2$). However, when looking at Figure \ref{fig:cluster_masses}, we see a large number of out-of-tail objects which pass these thresholds, indicating we
require tighter restraints on our cuts. This leads to fewer objects
which meet our SCC criterion, but improves the statistics on our background level estimate, $N^\text{SCC}_\text{out}/A_\text{out}$.

\begin{table*}
	\begin{center}
		\begin{tabular}{c c c c c c c c}
			Tail & $N^\text{SCC}_\text{in}$ & $N^\text{SCC}_\text{out}$  & $A_\text{in}$ & $A_\text{out}$ & 
			$N^\text{SCC}_\text{in}/A_\text{in}$ &  $N^\text{SCC}_\text{out}/A_\text{out}$ & $\Sigma_\text{SCC}$ \\
			& & & (kpc$^{2}$) &  (kpc$^{2}$) & (kpc$^{-2}$) & (kpc$^{-2}$) & (kpc$^{-2}$)  \\ \hline \hline
			West & 31 & 14 & 301.38 & 2855.21 & 0.103 & 0.0049 & $0.098 \pm 0.019$ \\
			East & 19 & 15 & 305.54 & 3019.22 & 0.061 & 0.0050 & $0.056 \pm 0.014$ \\
			\hline
		\end{tabular}
	\end{center}
	\caption{SCC counts for those detected within and outside the tail region. SCC excess ($\Sigma_\text{SCC}$)
		is defined as the difference betwen $N^\text{SCC}_\text{in}/A_\text{in}$ and
		$N^\text{SCC}_\text{out}/A_\text{out}$. Error bars of 1$\sigma$ from Poisson statistics are shown.
		SCC excess in the Western tail exceeds 5$\sigma$, while the Eastern tail exceeds
		4$\sigma$, showing statistically significant values of $\Sigma_\text{SCC}$ in both tails.}
	\label{table:excess}
\end{table*}

\begin{figure}
	\centering
	\includegraphics[width=1\linewidth]{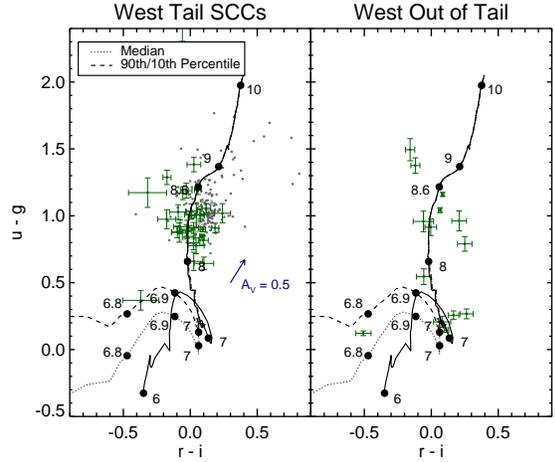}
	\vspace{-0.5cm}
	\caption{Star cluster candidates (SCCs) for the Western tail, as defined by colour cuts of $g - i$ $\leq 0.5$ and
		$11.5 \leq M_g \leq 9.25$. Gray points correspond to diffuse light. For comparison, we show 
		colours of objects outside the tail region on the right. Nebular tracks with emission from H$\beta$,
		H$\alpha$, [O \textsc{iii}], [O \textsc{ii}], and continuum emission are included. Dotted lines indicate median values of [O \textsc{iii}]/H$\beta$ and [O \textsc{ii}]/H$\beta$ emission, while dashed lines indicate the 90th
		and 10th percentile of [O \textsc{iii}]/H$\beta$ and [O \textsc{ii}]/H$\beta$ emission, respectively, from the KISS galaxy sample. 31 SCCs are detected in the Western tail, with a
		median age of 8.15 log yrs.}
	\label{fig:W_SCC_colour}
	\vspace{0.5 cm}
\end{figure}

\begin{figure}
	\centering
	\includegraphics[width=1\linewidth]{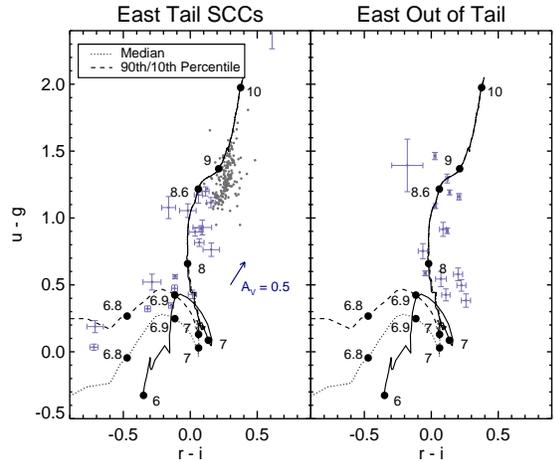}
	\vspace{-0.5cm}
	\caption{Same as Figure \ref{fig:W_SCC_colour} but for the Eastern tail. Several young objects show strong nebular emission. 19 SCCs are detected in the Eastern tail, with a median age of 7.96 log 
		yrs.}
	\label{fig:E_SCC_colour}
	\vspace{0.5 cm}
\end{figure}

A colour-colour diagram with only
these SCCs is shown in Figures \ref{fig:W_SCC_colour} and \ref{fig:E_SCC_colour}. Overlaid in gray are data points for the diffuse light of the respective tidal tail. The SSP models from \cite{marigo_08} do not include nebular emission. However, the nebular continuum, as well as
emission lines from H$\alpha$, H$\beta$, [O \textsc{iii}] and [O \textsc{ii}], can have strong effects on our colours
for young objects, with ages < 10 Myr. We use \texttt{Starburst99} (\citealp{SB99}) to include a nebular continuum,
as well as emission from H$\alpha$ and H$\beta$. Following \cite{sarah_10} and \cite{fedotov_11}, we find the strengths of the [O \textsc{iii}] and [O \textsc{ii}] lines
from the KISS sample of nearby low-mass star-forming galaxies (e.g., \citealp{salzer_05}). We use the median ratios of [O \textsc{iii}]/H$\beta$ and [O \textsc{ii}]/H$\beta$,
listed at 0.08 and 0.56, respectively. H$\alpha$ emission falls in the \textit{r} filter, causing the $r - i$ colour
to appear bluer. Both H$\beta$ and [O \textsc{iii}] fall in the \textit{g}-band, while [O \textsc{ii}] is in the 
\textit{u}-band. Depending on the relative strengths of the oxygen lines, this can cause the $u - g$ colour to become
bluer or redder. To show a possible range of nebular
emission, we also include a track setting [O \textsc{iii}]/H$\beta$ to its 90th percentile value (0.66) and [O \textsc{ii}]/H$\beta$
to its 10th percentile value (0.22). This provides a better fit for several data points in the Eastern tail, although the position of these clusters on the diagram is likely due
to a combination of varying emission line strengths and dust extinction. 

In examining Figure \ref{fig:W_SCC_colour}, we see a clear overdensity of SCCs in the Western tail compared to objects outside of the tail. In the Eastern tail (\ref{fig:E_SCC_colour}), a number
of SCCs exist at blue values of $u - g$ and $r - i$ which are distinct from those outside the tail. SCCs
in the Western tail overlap with the diffuse light (though skewed to the blue end), while those in the Eastern tail are distinct from the
diffuse light in the host tail. To quantitatively test this, we apply the Kolmogorov--Smirnov (KS) test to the colour distribution of the diffuse light and SCCs. This test allows us to determine the probability that two samples 
were drawn from independent distributions. A \textit{p}-value of less than 0.013 indicates the populations are distinct from one another at more than a 2.5$\sigma$ confidence level. For the $u - g$ colours, we find \textit{p}-values
of 0.028 and $4.2 \times 10^{-11}$ for the Western and Eastern tails, respectively. For the $r - i$
colours, we have \textit{p} values of $3.9 \times 10^{-5}$ and $8.4 \times 10^{-13}$, for the Western and Eastern tails. These results
show the diffuse light and SCC colour distributions are distinct from each other in both tails.


\subsubsection{Cluster Masses and Ages} \label{sec:cluster_mass}

\begin{figure}
	\centering
	\includegraphics[width=1\linewidth]{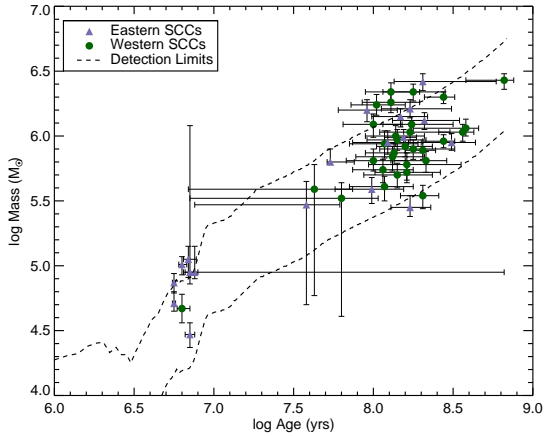}
	\vspace{-0.7cm}
	\caption{Ages and masses of SCCs for the Western (dark green) and Eastern (purple) tail. Dashed lines indicate our
		SCC detection threshold, as determined by our colour and magnitude limits. Two detection limits are formed from our upper and lower magnitude cuts. The gap between the young
		objects in the Eastern tail and the main distribution at ages of 8.0 - 8.8 log yrs shows we are seeing
		an isolated burst of star formation.}
	\label{fig:mass_age}
	
\end{figure}

Ages and masses of our sources are found using the 3DEF method (\citealp{bik_03}). This uses a maximum likelihood
estimator to find the ages and masses of each cluster, by adjusting the colour excess $E(B-V)$ to fit the observed
colour in each filter to a given SSP evolutionary model. Results for the total masses and median ages of SCCs are 
shown in Table \ref{table:mass_all}, with masses and ages shown in Figure \ref{fig:mass_age}; dashed lines represent
our SCC detection limits based on our colour and magnitude criteria from Section \ref{sec:cluster_colours}. Error bars 
show the maximum and minimum age and mass for each data point. At low masses,
there are several points which do not fall within the detection band. These objects are subject to internal extinction,
which dims them to fall within our SCC colour and magnitude cutoffs. There is a gap in age between the young objects in the Eastern
tail and the main distribution with ages of 8.0 - 8.8 log yrs. This indicates a recent small burst of isolated star formation in the Eastern tail, small enough that only low mass clusters are present.  Other similar bursts could have occurred in either tail between 7.0 and 8.0 log yrs, but would have faded from view by the present.  It is clear, however that there is not continuous star formation at the level of that seen during the main interaction period, 8.0 - 8.8 log yrs ago. 



\subsection{Spatial Distribution of SCCs and Diffuse Light}

\begin{figure*}
	\centering
	\subfigure{
		\centering
		\includegraphics[width=0.45\linewidth]{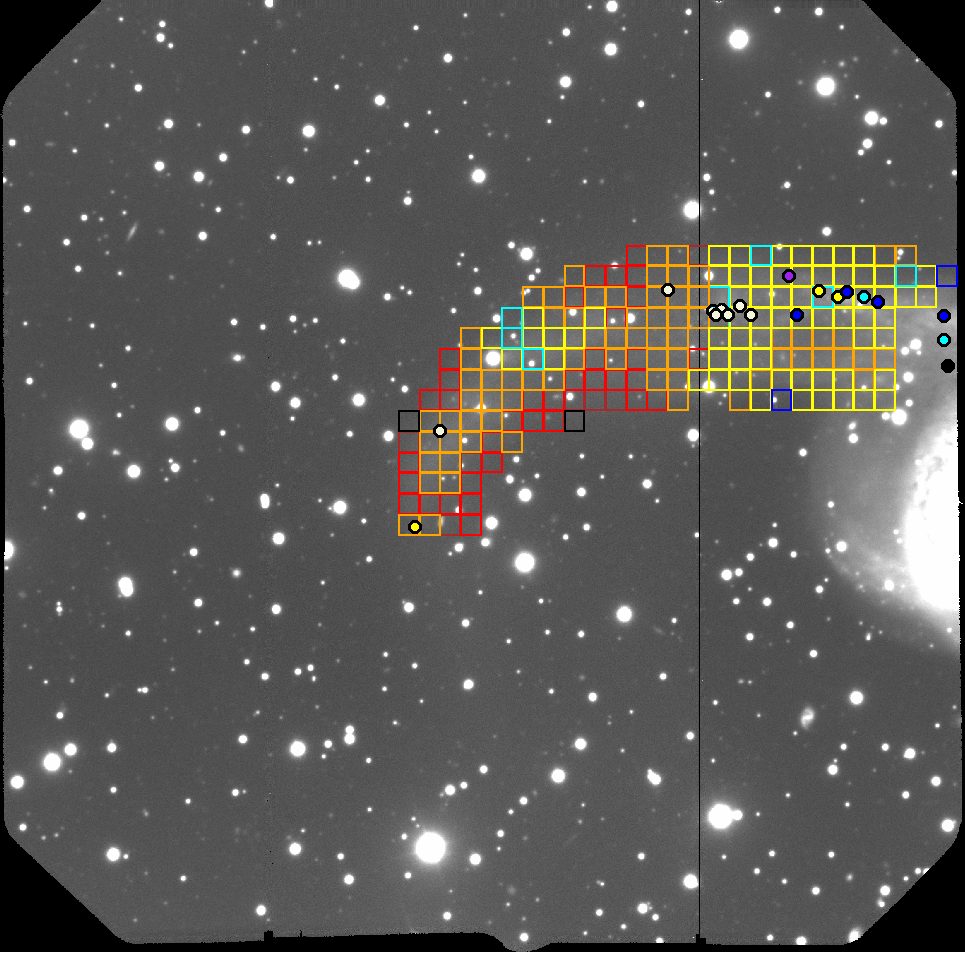}
	}
	\hspace{1 cm}
	\subfigure{
		\includegraphics[width=0.45\linewidth]{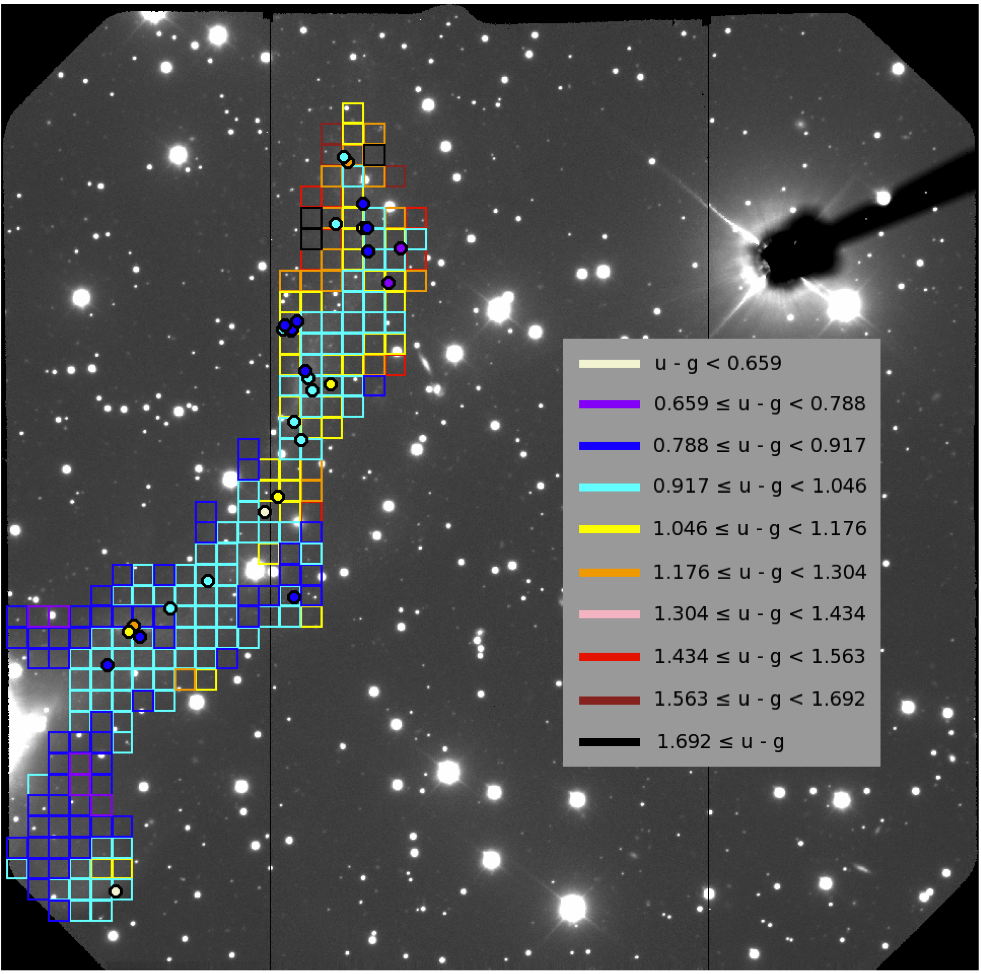}
	}
	\caption{\textit{Left}: Spatial distribution of diffuse light and SCC $u - g$ colours for the Eastern tail. \textit{Right}: Same, but for the Western tail. Colour
		code is reproduced from Figure \ref{fig:cluster_masses}. Each tail shows a gradient
		in colour across their length,
		suggesting a decrease in the fraction of young stars within the diffuse light along the tail
		from centre to tip.}
	\label{fig:colour_tail_SCCs}
	
\end{figure*}

Spatial maps of tail colours and SCCs are shown in Figure \ref{fig:colour_tail_SCCs} for both tails. Measurement boxes from
Figures \ref{fig:Boxes} are colour coded to indicate their $u - g$ colours. SCC positions and $u - g$ 
values are added in a similar manner. 

The difference in age between the tails is evident in
the abundance of younger regions, represented by cyan and yellow, in the Western tail, compared to older regions, represented by orange and red, in the Eastern tail. The spatial location of SCCs does not appear to influence the colour of the diffuse light. This is not too surprising, as these
objects were masked out in Section \ref{sec:masking}. 

Both of the tails show colour gradients across their lengths, with bluer colours near the galactic bulges
and redder colours at the far tips. Likewise, the edges of the tails appear redder than their interiors. Such
gradients were also seen among several tails in \cite{mulia_15}.

\section{Discussion}
\label{sec:5}
\begin{figure}
	\centering
	\includegraphics[width=1\linewidth]{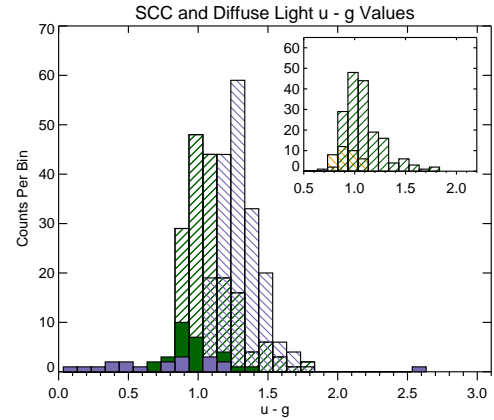}
	\vspace{-0.7cm}
	\caption{$u - g$ colour distribution for SCCs and diffuse light in NGC 3256W (dark green) and E (purple), with bin
		size of 0.1. Hashed histograms indicate diffuse light measurements, while solid histograms represent the SCCs
        Colour distributions for the separate diffuse structure in the Western tail are shown in the inset graph in orange, plotted alongside
        those of the Western tail itself (dark green). The diffuse light colours between the two tails
        are clearly distinct from one another. The distribution of SCC colour in the Western tail
        overlaps its diffuse light colour distribution, suggesting a common origin. SCCs in the Eastern
        tail are much younger than any of the SCCs in the Western tail.}
	\label{fig:colour_hist}
\end{figure}

Distributions of $u - g$ colours for diffuse light and SCCs are shown in Figure \ref{fig:colour_hist}.
We can summarize Figure \ref{fig:colour_hist} with the following points:
\begin{enumerate}
	\item The difference in $u - g$ colour between the Eastern and Western tail shows
	the total diffuse stellar light of the two tails are separated in age by over 500 Myr,
	with a significantly larger contribution from a young population in the Western tail.
	\item Assuming the two tails were formed at the same time, we find the stellar masses of the
	Eastern and Western tails to be dominated by old ($\sim$10 Gyr) populations drawn from the parent galaxies.
	\item While the Western tail contains a significant old ($\sim$10 Gyr) population, the stellar light is largely
	comprised of a population which formed soon after the interaction, with an age of 8.29 
	log yrs.
	\item The similarity of the $u - g$ colour distribution of SCCs in the Western tail compared
	to the diffuse light gives credence to the idea that the Western tail is comprised of disrupted star
	clusters formed shortly after the formation of the tail, with the ones we see today being the ones that survived.
	\item The stark contrast in $u - g$ colours for SCCs in the Eastern tail as compared to the diffuse
	light colours suggests some very recent star formation as compared to the age of the tail.
	
\end{enumerate}
Stellar populations in the two tails of NGC 3256 have different compositions. We base our conclusions on the assumption that the
tails were formed at the same time. It's possible to interpret Figure \ref{fig:LSBphot} as due to the Eastern tail having formed
first, with the stellar light we observe being formed from a burst of star formation caused by that earlier interaction. 
However, prograde collisions between 
galaxies are required to form tidal tails (\citealp{donghia_10}; \citealp{toomre_72}), as in the case of NGC 3256. In such a case, these tails form simultaneously, which
indicates a common
age between NGC 3256's tails. \cite{lipari_00} found three nuclei within the centre of NGC 3256, leading them to 
construct a merger scenario involving two separate mergers and three galaxies. The Eastern tail could be explained
to have formed during the first merger, while the Western tail formed later with a prograde interaction between the
merger remnant and the third galaxy.
However, \cite{lipari_00} suggests a more likely scenario with the two major galaxies initially merging, and a smaller satellite 
galaxy merging after the major interaction, supported in further work by \cite{lipari_04}. Additionally, \cite{english_03}
only observed two broad H\textsc{i} tails, leading them to conclude that NGC 3256 is likely created by a two galaxy prograde
merger. Given these works, it is very probable that the two tails were formed at the same time. We also find possible evidence for the
small satellite merger in the separate diffuse structure, seen in the Western tail. This structure is younger than either tail, supporting
the theory of a late minor merger, occurring after the major merger.

The diffuse light in the Western tail is dominated by a younger population, formed soon after the formation of the tail. This population is also present
in the Eastern tail, although at a lower concentration. The source of the young population is uncertain, as stars may form in a variety of
methods: in bound clusters, unbound stellar associations, down to near individual stellar formation. The fraction of stars forming in clusters
can be up to 70\% in regions of large gas density (\citealp{kruij_12_2}). However, these clusters are subject to frequent tidal shocks which will preferentially destroy low mass clusters (\citealp{kruij_12}), dispersing their material into the diffuse tidal light and making clusters an
ideal candidate for the source of our young population. The destruction of star
clusters is modeled in two parts: number loss (removal of stars in a cluster) and mass loss (removal of mass in a cluster). 
Number loss can be understood as effects from stellar feedback, such as stellar winds and supernovae, which expel gas in
a cluster, causing it to become unbound. Mass loss will remove stars from a cluster via two-body interactions. Number loss,
also known as infant mortality, will only be important for the first $\sim$10 Myr, as the hot and massive stars evolve and explode. The remnants of
these clusters are thus seen as the diffuse light of the tail. A similar effect is seen in NGC 7714 -- ages of H\textsc{ii} regions within
the tidal tails have been shown to be older than star clusters residing within them, suggesting previously formed clusters have been dispersed
and surround the newly conceived clusters (\citealp{peterson_09}). The coincidence of peaks in histograms of $u - g$ colour between the diffuse light of the
Western tail and its SCCs (Figure \ref{fig:colour_hist}) strongly suggests the two are intertwined. Results of KS tests
show the populations are distinct from each other, however, this is likely due to the presence of an old stellar population in the diffuse light of the tail. The timescale between
tail formation and cluster formation is separated by $\sim$200 Myr. We compare this to simulations of NGC 4038/9, a similar merger,
which found a peak in star formation $\sim$25 Myr after the interaction (\citealp{renaud_15}). 

Individual masses of SCCs in the Western tail are on average more massive than in the Eastern tail, as shown in Figure \ref{fig:mass_age}, although
the Eastern tail has several very young, low mass objects absent in the Western tail. The greater number of
SCCs for the Western tail can be attributed to its larger H\textsc{i} mass. This can be related back to the diffuse light as well, as the
Western tail's higher abundance of gas led it to form more star clusters which could be disrupted and dispersed in the tail. We compare
the ages of our clusters to those found in the nuclei from previous studies. \cite{trancho_07_2} spectroscopically studied 23 star clusters
inside the centres of the galaxies, finding an average age of $\sim$10 Myr. \cite{zepf_99} performed photometry on several hundred objects
within the centre finding bright and blue objects. Our SCCs are substantially older than those found within the nuclei, suggesting the 
star formation in the tails was cut off earlier relative to the interior. This is consistent with spectroscopic observations of  star
clusters in the Western tail by \cite{trancho_07_1} and photometric analysis of SCCs in the Eastern tail by \cite{mulia_15}, both of 
whom found ages of clusters to be older than those in the interior. A similar viewpoint is shared in the Antennae simulations, which
show a cessation of star formation in the tails, while star formation in the interiors is ongoing (\citealp{renaud_15}). Star formation in
the interiors can be stimulated as material previously thrown out during the initial encounters falls back into the centre of the potential
well.

Of particular interest are the very young objects found in the Eastern tails, with $u - g < 0.7$. These objects indicate relatively recent
star formation in a tail whose diffuse light suggests a limited star formation history. It is possible these objects are now being formed as
material falls back through the tail to the interiors, creating turbulence in the H\textsc{i} gas and sparking small bursts of star formation. We do not see these very young objects in the
Western tail. However, note that they are low mass, with masses $< 10^5 M_{\odot}$. They will fade from detectability as they age (see Figure \ref{fig:mass_age}); additionally, they may subject to tidal shocks from regions of dense gas which can disrupt them. This suggests that there may be small star
formation events periodically in both tails, but the evidence of these events will rapidly fade away.

\begin{figure*}
	\centering
	\subfigure{
		\includegraphics[width=0.4\linewidth]{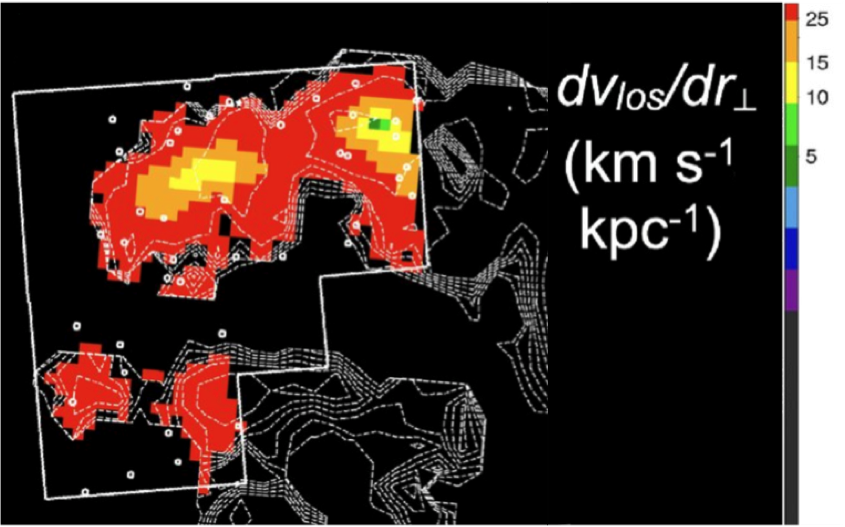}
	}
	\hspace{0.3 cm}
	\subfigure{
		\includegraphics[width=0.4\linewidth]{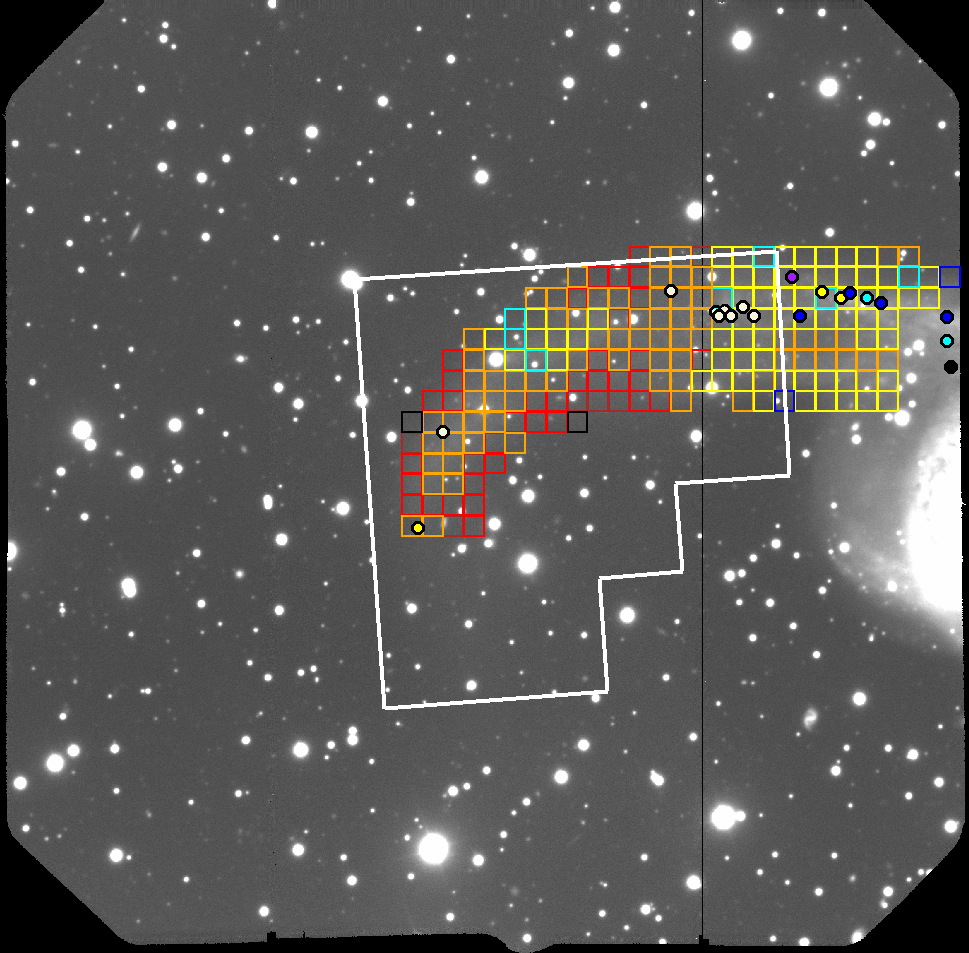}
	}
	\caption{\textit{Left}: H\textsc{i} shear measurements for the Eastern tail (\protect\citealp{mullan_13}). The WFPC2 footprint
		is overlaid in solid white. Dashed white contour lines correspond to log $N_{\mathrm{H\textsc{i}}}= 20.0 - 21.4$ cm$^{-2}$
		in steps of 0.2 dex, while white circles are SCCs as determined by \protect\cite{knierman_03}. \textit{Right}: $u - g$ spatial
		map of the Eastern tail, with the WFPC2 footprint overlaid in white. Two regions of relatively low shear can be seen (yellow and orange regions in the left-hand panel), one
		at the edge of the WFPC2 footprint near the bulge, the other in the centre. The first can be matched in our $u - g$ spatial
		distribution to the group of young SCCs mentioned in the previous paragraph. The second corresponds to a patch of
		yellow boxes of diffuse light (right-hand panel).}
	\label{fig:E_shear}
\end{figure*}

\begin{figure*}
	\centering
	\subfigure{
		\includegraphics[width=0.4\linewidth]{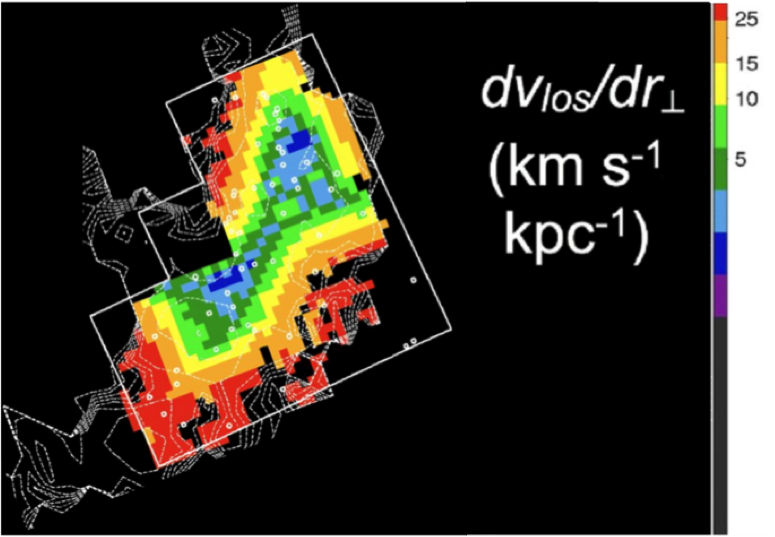}
	}
	\hspace{0.3 cm}
	\subfigure{
		\includegraphics[width=0.4\linewidth]{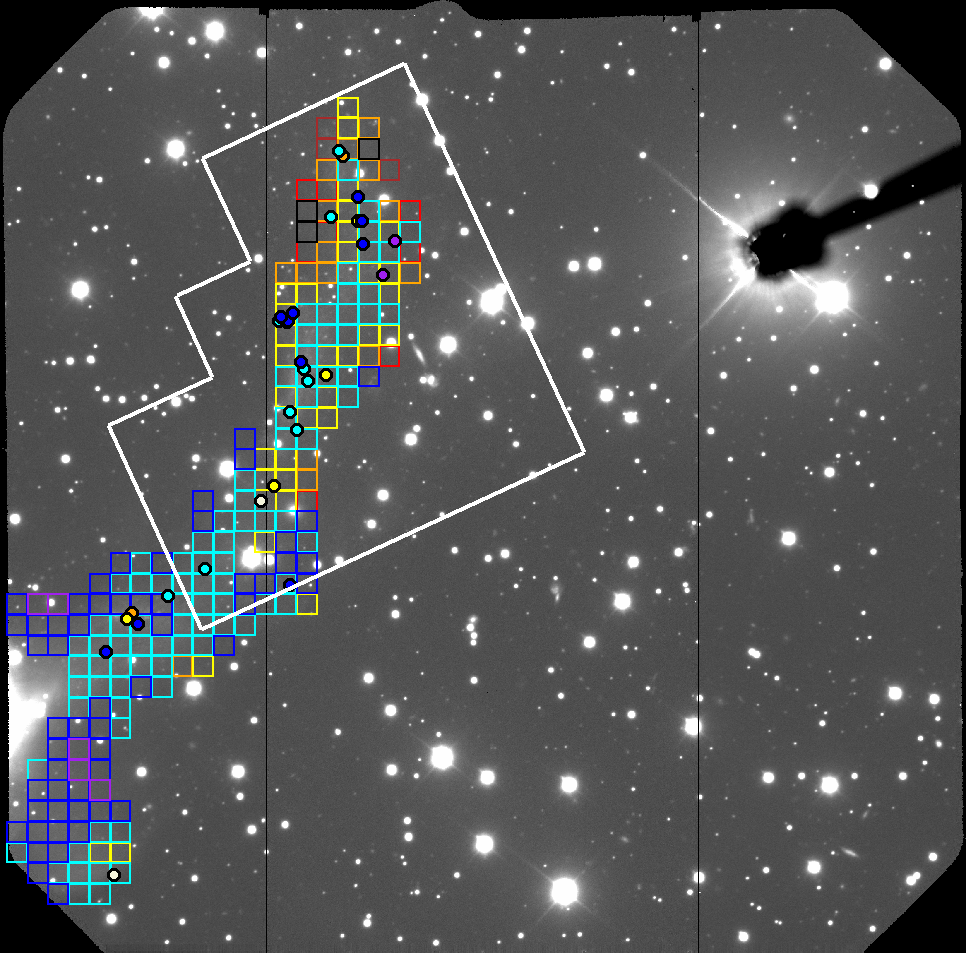}
	}
	\caption{Same as Figure \ref{fig:E_shear}, but for the Western tail. The majority of detected SCCs within the WFPC2 footprint
		lie in the spine of low shear (green and blue in the left-hand panel), with measurements less than 5 km/s/kpc.}
	\label{fig:W_shear}
\end{figure*}

\cite{mullan_13} found that tidal tails with large H\textsc{i} line of sight velocity dispersion $\sigma_{\mathrm{los}}$ and high
H\textsc{i} column densities were ideal locations for SCCs. Both the 3256 tails fit these criteron, and the majority of their SCCs, as
determined by \cite{knierman_03} using \textit{HST} WFPC2 data, lie in H\textsc{i} pixels with these characteristics. However, the tails can be distinguished by measurements of shear ($\mathrm{d}v_{\mathrm{los}}/\mathrm{d}r_\bot$), with the Western tail's shear about one-third
that of the Eastern's. \cite{mullan_13} found that SCCs were preferentially located in regions of low shear, which is suggestive
that these areas can be ideal for star formation. We reproduce spatial maps of H\textsc{i} shear from Figures 4.18 and
4.19 in \cite{mullan_13}, and compare them to our spatial $u - g$ maps in Figure \ref{fig:E_shear} for the Eastern tail and
Figure \ref{fig:W_shear} for the Western tail.

From Figure \ref{fig:E_shear}, two pockets of low shear are visible in yellow. The first of these, closest to the bulge, corresponds to the location of the youngest SCCs in the Eastern
tail, visualized as beige circles. The second low shear region does not contain any SCCs, although the diffuse light near this area is younger than its
surroundings.

The Western tail has a ridge of low shear seen in Figure \ref{fig:W_shear}, running from the middle of the tail to the tip. We find the majority of SCCs within the
H\textsc{i} field of view reside in this region, with few ($\sim$3 out of 23) outside. The diffuse light appears to follow the shear
as well. At the tip of the tail, the low H\textsc{i} shear seen in blue matches the yellow diffuse light boxes. As the shear
increases to yellow and orange, the diffuse light reddens and the boxes shift to orange and red.

The effect of dust on the diffuse light in these tails is similar to that of an older population; in either case the colours are reddened. To
investigate this effect, we redden the Eastern tail colour to match the Western's, by minimizing
the distance between the two values. We find the amount of extinction needed for this is $A_V \approx 0.6$,
which we find unrealistic, as tidal tails are regions of relatively low extinction (\citealp{tran_03}; \citealp{bastian_05}). Additionally,
examination
of archival \textit{GALEX} NUV data (\citealp{paz_07}) show similar brightnesses for the two tails, suggesting
similar levels of low extinction. Extinction in the Eastern tail would dim the \textit{GALEX} data, with respect to the Western
tail, but this is not seen. \cite{mulia_15} examined the $B - V$ colour distribution within the Eastern tail, finding
negligible levels of reddening, supporting the lack of dust within the tails.

We finally look at extinction in the three star clusters spectroscopically studied in the Western tail from \citealp{trancho_07_1},
which are listed at 0.0, 0.3, 0.5 $A_V$ mag, showing minimal extinction. Additionally, as mentioned in Section \ref{sec:cluster_mass}, several of our
young SCCs are subject to internal extinction. Extinction values of these objects range from 1.1 to 0.7 $A_V$ mag. Dust can be expected
to associate with star clusters, as dust is needed for star formation, so we expect to see the most extinction in these regions. However, 
such levels of extinction exist for a small handful of objects, and as these are masked and do not factor into our LSB measurements, we are lead to believe extinction does not significantly affect our analysis of the stellar age and mass distribution.

\section{Conclusion}
\label{sec:6}
Our observational program has proven successful in allowing us to characterize the stellar populations of our tidal tails.
We find NGC 3256W to be bluer than its twin tail, NGC 3256E. Measured colours indicate that diffuse light in the Western tail
has a large contribution from a young population formed after the interaction, perhaps from dispersed star clusters, as compared to
the Eastern tail, which is primarily illuminated by an old population derived from the host galaxy. 
Both tails exhibit colour gradients along their lengths, suggesting
a gradient in the time scale of star formation. Despite these colour differences, both tails appear to be dominated in mass by
an old, underlying population, originating from the interacting galaxies.



Analysis of SCCs shows a lack of old objects in either tail (> $10^9$ yr), but a clustering of objects below 400 Myr in 
the Western tail and Eastern tail. The Eastern tail shows an interesting clustering of young objects, with ages < $10^7$ yr. These objects
are low mass structures and are not likely to be detected as they age, disappearing as they fade beyond our detection limits, or are dispersed into the tail. The $u - g$ colour distribution of the
Western SCCs is proven to be distinct from the diffuse stellar light in the Western tail through KS probability tests, however
the peaks of the colour distributions of the diffuse light and SCCs match well, suggesting these objects and regions are intertwined. The
fact that the KS test shows separate populations can be explained by the addition of an old, underlying stellar population to
the diffuse light.

NGC 3256 has been shown in past studies to contain a large number of SCCs compared to other systems (\citealp{mullan_11}; \citealp{knierman_03}).
We plan to apply our current observational program to additional tidal tail systems with varying amounts of SCCs and H\textsc{i} properties,
particularly those with low numbers of SCCs. The presence of a tidal dwarf galaxy may also play a role in determining the composition of tidal tail diffuse light. \textit{GALEX} observations
of the Antennae galaxy reveal a gradient in colour along the tail, with bluer colours near the tip of the Southern tail, where two tidal dwarf galaxies
reside (\citealp{hibbard_05}). \cite{knierman_03} found that tails with tidal dwarfs did not contain as many SCCs as those without them.
Have these structures already dispersed their clusters into their tails, or has there been a complete absence of star cluster formation? This remains to be seen.

We would like to thank the anonymous referee for helpful comments which have improved the quality and content of this paper.
Based on observations obtained at the Gemini Observatory (program ID GS-2013A-Q-57, processed using the Gemini IRAF package), which is operated by the Association of Universities for Research in Astronomy, Inc., 
under a cooperative agreement with the NSF on behalf of the Gemini partnership: the National Science Foundation (United States), the National 
Research Council (Canada), CONICYT (Chile), Ministerio de Ciencia, Tecnolog\'{i}a e Innovaci\'{o}n Productiva (Argentina), and Minist\'{e}rio da 
Ci\^{e}ncia, Tecnologia e Inova\c{c}\~{a}o (Brazil). 
The Digitized Sky Surveys were produced at the Space Telescope Science Institute under U.S. Government grant NAG W-2166. The images of these surveys are based on photographic data obtained using the Oschin Schmidt Telescope on Palomar Mountain and the UK Schmidt Telescope. The plates were processed into the present compressed digital form with the permission of these institutions. 
SCG thanks the Natural Science and Engineering Research Council of Canada for support.
KK is supported by an NSF Astronomy and Astrophysics Postdoctoral Fellowship under award AST-1501294.
The Institute for Gravitation and the Cosmos is supported by the Eberly College of
Science and the Office of the Senior Vice President for Research at
the Pennsylvania State University.

\bsp	
\label{lastpage}
\end{document}